\documentclass[A4,12pt]{extarticle}
\usepackage{setspace}
\usepackage{amsmath}
\usepackage[top=1in,bottom=1in,right=1in,left=1in]{geometry}
\usepackage{caption}
\usepackage{comment}
\usepackage{psfrag}
\usepackage{graphicx}
\usepackage{hyperref}
\usepackage{subfigure}
\usepackage{authblk}
\usepackage{verbatim}
\usepackage{multicol}
\usepackage{color,soul}
\usepackage{xcolor}
\usepackage{bibcheck}
\usepackage[natbib=true,sorting=none,style=nature]{biblatex}
\title{Charged AdS Black Holes in Presence of String Cloud and Cardy-Verlinde Formula}
\author{Rishi Pokhrel\thanks{E-Mail: rishipokhrel.smit@gmail.com and rishi\_20211037@smit.smu.edu.in} }

\author{Tanay K. Dey\thanks{E-mail: tanay.dey@gmail.com and tanay.d@smit.smu.edu.in}}
\affil{Department of Physics, Sikkim Manipal Institute of Technology, Sikkim Manipal University, Majitar, Rangpo, East Sikkim, 737136, India.}

\date{}
\begin{document}
	\maketitle

	\begin{abstract}
		In this study, charged AdS black holes with a string cloud are taken into consideration. We identify three distinct black hole solutions within a specific temperature range, which subsequently merge into a single black hole beyond this range in the case of spherical space with low chemical potential and string density.  However, in the case of large chemical potential and string density, only one black hole solution exists for all temperatures. Notably, we observe that for flat and hyperbolic space, there is only one black hole solution irrespective of the chemical potential and string density. By subtracting the contribution of the extremal black hole from the on-shell Euclidean action, the thermodynamical quantities associated with these black holes are calculated.  We also investigate the stability of these black holes and find that medium-sized black hole is unstable. In the dual boundary theory side we verify the existence of bound state of quark-antiquark pair. Finally, we establish the Cardy-Verlinde formula for the boundary theory and we observed an additional energy term arising because of the presence of string cloud.
	\end{abstract}

\clearpage


\section{Introduction}
The AdS/CFT correspondence \cite{Maldacena1999,Witten1998, Witten2014} describes a duality between strongly coupled $\mathcal N = 4$, $SU(N)$ gauge theories on the boundary of Anti-de Sitter spacetime and weakly coupled gravity theories in one higher dimension in the bulk of AdS spacetime. This duality allows for the study of strongly coupled gauge theories, which would otherwise be difficult to study. The thermodynamics of the weakly coupled gravity theory can be mapped onto the strongly coupled boundary $\mathcal N = 4$, $SU(N)$ gauge theories. The Hawking-Page phase transition \cite{Hawking1983} in the weakly coupled gravity theory corresponds to the confined and deconfined phase transition in the strongly coupled gauge theory \cite{Witten2014}. Furthermore, in \cite{Verlinde2000a} Verlinde showed that the entropy of the dual gauge theory can be expressed as Cardy like formula for two dimensional CFT \cite{Cardy:1986ie}. Since then, this is commonly known as Cardy-Verlinde formula. Subsequently, there has been several works reported in the literature to establish a similar formula for the dual CFT theories for different types of corresponding AdS black hole solutions \cite{ Savonije2001,Cai2002b,Cai2001c,Brevik2004,Biswas2004,Biswas2001a,Cappiello2001, Nojiri2002,Nojiri2001d,Cai2004}.\\

In $\mathcal N = 4$ $SU(N)$ gauge theory, quarks with fundamental representation are not present. However, it is possible to incorporate these quarks by adding the spacetime filling $D7$-brane in the bulk spacetime. In \cite{Karch2002,Bigazzi:2011it} authors consider limited number of filling branes and in\cite{Headrick:2007ca,Kumar:2012ui} the authors considered a large number of filling branes in AdS spacetime. The fundamental quarks in the gauge theory come from the end points of the strings stretching from the $D7$-brane to the horizon of the bulk spacetime. The body of the strings represent the gluonic field. Several studies have explored the coupling of the string cloud with gravity theory \cite{Chakrabortty2011a,Dey2018,Herscovich2010a,Ghaffarnejad2018a,Ghanaatian2019a,Dey:2020yzl,Dey:2023inw}. In \cite{Chabab2020,Liang2021,Yin2021a}, charged AdS black hole solutions surrounded by quintessence in presence of string cloud are considered in spherical spacetime.\\

This study examines charged AdS black hole solutions in the presence of a string cloud for spherical, flat, and hyperbolic spaces in general dimensions. It is observed that, regardless of the spacetime curvature, string density, and chemical potential, an extremal black hole can exist at zero temperature. The thermodynamical quantities related to the black hole solutions have been computed from the Euclidean action by subtracting the contribution of the extremal black hole. The analysis of the thermodynamic quantities demonstrates that the first law of thermodynamics remains valid for these black hole solutions and provides the number of possible black hole solutions and their nature for a given temperature. For spherical curvature and small chemical potential a small black hole exists at low temperature and above a critical temperature, two more black holes nucleate. Depending on their size, these black holes are categorized as small, medium, and large black holes. Within a particular temperature range, these three black holes coexist. At the upper limit of the temperature range, the medium and small black holes merge and disappear, while only the large one remains above the temperature range. The small and large black holes have positive specific heat, while the medium black hole has negative specific heat. Consequently, the first two black holes are thermodynamically stable, while the latter one is unstable. Additionally, we calculate the free energy of these black holes and demonstrate that, at low temperatures, the small black hole is thermodynamically favoured, while the large black hole is preferred at high temperatures. At a critical temperature, a phase transition similar to Hawking and Page between the small and large black hole is possible, corresponding to the confined and deconfined phase transition in the dual boundary gauge theory. For spherically curved space with large chemical potential and string density, only one black hole exists, which is also true for flat and hyperbolic space. Therefore, AdS space is not a valid configuration in the system, indicating that there may not be any confined state of quark-antiquark in the corresponding dual gauge theory. As a preliminary step to investigate the gauge theory, we examine the distance between quark-antiquark in the boundary for the small and large black hole solution of the corresponding bulk theory. It has been observed that for the small black hole spacetime the connecting string between quark and antiquark does not break even for large distance whereas the connecting string breaks down after certain distance of quark-antiquark pair for large black hole spacetime. Therefore we conclude that though in the bulk theory AdS spacetime no longer exist but in the dual gauge theory bound state of quark-antiquark pair is a valid state at low temperature. For further confirmation, we examine whether the Cardy-Verlinde formula is valid in this system or not. Ultimately, we establish the Cardy-Verlinde formula for the boundary gauge theory through two distinct methods. The first approach derives the formula from the Casimir energy of the boundary theory, while in the second approach the formula arises naturally from the Landau function of the boundary gauge theory. In the Cardy-Verlinde formula, we observed an additional energy term arising because of the presence of string cloud.  It is observed that the formula may not be valid for hyperbolic curvature space. \\

 The work is organized as follows: in section (\ref{section_action_eom_black_hole_solution}) we first set up the action with the addition of the string cloud into the system and then constructed the Einstein equation of motion for the system. From the equation of motion, black hole solutions are found. In order to study the thermodynamics of the black holes, we compute the Euclidean action in section (\ref{subsection_temperature}) and then calculate entropy, energy, specific heat, free energy and Landau function. Subsequently, we find the number of black hole solutions that can be possible in a particular temperature and their stability in section (\ref{subsection_black_hole_phases}).  We study the existence of bound state of quark-antiquark pair in section (\ref{section_quark_antiquark_distance}) and then in section (\ref{cardyverlinde}), the Cardy-Verlinde formula has been obtained for the dual boundary gauge theory corresponding  to the charged AdS black hole with string cloud.  Finally, in section (\ref{summary}), a summary and discussion have been presented.


	\section{Black Hole Solution}
	\label{section_action_eom_black_hole_solution}
	After the brief introduction in the previous section, this section discusses the action, equation of motion, and black hole solutions for charged AdS black holes in the presence of string cloud. The string cloud represents the introduction of string in the background of charged AdS black holes. These strings are open and extend from the AdS space boundary to the black hole horizon. According to the holographic principle, this system represents a specific state in the gauge theory in the presence of quarks. Following \cite{Chakrabortty2011a,Dey2018}, we consider the action of charged AdS black holes in presence of string cloud.\\

\subsection{Action}
To investigate the $(n+1)$-dimensional bulk theory resulting from the charged AdS spacetime with a string cloud, we consider the following action\footnote{To write the action we follow \cite{Chamblin1999a}, where they have constructed the five dimensional action on the $AdS_5 \times S^5$ from the ten dimensional type IIB supergravity. The charge of the black hole is introduced from the spinning of the $S^5$ sphere. Then they have extended the action for arbitrary dimension. Due to the dimensional reduction the higher order correction terms also arise but we are not considering those terms here. In our system, we have added the contribution of the string with the action of \cite{Chamblin1999a}.}

	
	\begin{equation}
		\label{action_total}
		I = -\frac{1}{16 \pi G_{n+1}}\int_M d^{n+1} x \sqrt{-g} [ R- 2 \Lambda - F^2] + I_{SC}.
	\end{equation}
	Here $G_{n+1}$ is the Gravitational constant, $g$ is the determinant of the spacetime metric tensor $g_{\mu \nu}$, $\Lambda$ represents the cosmological constant term, $F^2 = F_{\mu \nu}F^{\mu \nu}$ is the Maxwell charge term.  We have added the contribution of the string cloud as $I_{SC}$, where
	\begin{equation}
		\label{string_action}
		I_{SC} = -{\frac{1}{2}} \sum_i {\cal{T}}_i \int d^2\xi {\sqrt { |h_{\alpha\beta}|}}h^{\alpha \beta} \partial_\alpha X^\mu \partial_\beta X^\nu g_{\mu\nu}.
	\end{equation}
	Here ${\cal T}_i$ is the tension of the $i^{th}$ string and $h^{\alpha \beta}$ is the world-sheet metric. With these setups we now move on to set the equation of motion.

	\subsection{Equation of Motion}
	\label{subsection_eom}
		The equation of motion with respect to the spacetime metric $g_{\mu \nu} $  can be written as,
		\begin{equation}
			\label{equation_of_motion_wrt_g_munu}
			G_{\mu \nu} + \Lambda g_{\mu \nu} = T_{\mu \nu}^{EM} + 8 \pi G_{n+1}T_{\mu \nu}^{SC}.
		\end{equation}
Here
\begin{equation}
	\label{equation_einstein_tensor_G_munu_}
	G_{\mu \nu} = R_{\mu \nu} - \frac{1}{2} g_{\mu \nu} R.
\end{equation}
	$T_{\mu \nu}^{EM} $ is the electromagnetic energy momentum tensor which is given by
	\begin{equation}
		\label{electromagnetic_stress_energy_tensor}
		T_{\mu \nu}^{EM} = 2 {F_\mu}^\lambda F_{\nu \lambda} - \frac{1}{2} g_{\mu \nu} F^2
	\end{equation}
	and $T_{\mu \nu}^{SC} $ is the energy momentum tensor corresponding to the string cloud (SC) given by
	\begin{equation}
		\label{string_cloud_stress_energy_tensor}
		T^{(SC)\mu \nu} = -\sum_i {\cal{T}}_i \int d^2 \xi \frac{1}{\sqrt{|g_{\mu \nu}|}}\sqrt{|h_{\alpha \beta}|} h^{\alpha \beta} \partial_\alpha X^\mu \partial_\beta X^\nu \delta_{i}^{n+1}(x-X).
	\end{equation}
	In equation (\ref{string_cloud_stress_energy_tensor}), the delta function represents the source divergences due to the presence
of strings.\\
We assume that the strings are uniformly distributed over the $(n-1)$ spatial directions and write the density $a$ as,
	\begin{equation}
		\label{string_density_equation}
		a(x) = T\sum_i \delta_i^{(n-1)}(x-X_i).
	\end{equation}
	

\subsection{Solution}
\label{solution}
The equation of motion with respect to the electromagnetic field $A_\mu$ can be written as,
	\begin{equation}
		\label{maxwell_equation_of_motion}
		\nabla_\alpha F^{\alpha\beta}=0.
	\end{equation}
	For static and symmetric metric
	\begin{equation}
		\label{metric_ansatz}
	ds^2 = -V(r) dt^2 + \frac{dr^2}{V(r)} + r^2 d\Omega_{n-1}^2,
	\end{equation}
	where $d\Omega_{n-1}^2$ denotes the metric on unit $(n-1)$ sphere.
	The solution of equation (\ref{maxwell_equation_of_motion}) can be written as,
	\begin{equation}
		\label{maxwell_equation_of_motion_solution}
		F_{rt} = \sqrt{\frac{(n-1)(n-2)}{2}}\frac{q}{r^{n-1}}.
	\end{equation}
	The parameter $q$ is related with the charge of the black hole,
	\begin{equation}
		\label{charge_Q_value}
		Q = \sqrt{2 (n-1)(n-2)}\left(\frac{\omega_{n-1}}{8 \pi G_{n+1}}\right)q.
	\end{equation}
The chemical potential developed due to the presence of charge is
	\begin{equation}
		\label{electrostatic_potential_equation}
		\Phi = \sqrt{\frac{(n-1)}{2(n-2)}}\frac{q}{r_+^{n-2}}.
	\end{equation}
	With all these parameters, we now look for the solution of equation (\ref{equation_of_motion_wrt_g_munu}), considering the metric ansatz (\ref{metric_ansatz}).
	Under the static gauge condition $ X^t = \xi^0,\quad X^r = \xi^1, \quad X^\mu = y^\mu \quad \rm{for} \quad \mu \ne t \quad\rm{and}\quad \mu \ne r$, the non-vanishing components of  the string cloud energy momentum tensor $T^{(SC)\mu \nu}$ can be computed as,
	\begin{equation}
		\label{string_cloud_non_vanishing_components}
		T^{(SC)tt} = -\frac{a g^{tt}}{r^{n-1}},\,\,\,T^{(SC)rr} = -\frac{a g^{rr}}{r^{n-1}},
	\end{equation}
	\begin{equation}
		{T^{(SC)}}_t^t = {T^{(SC)}}_r^r=-\frac{a}{r^{n-1}}, \, \text{with $a > 0$}.
	\end{equation}
	Here, the density $a$ is assumed to be a constant. Following \cite{Chakrabortty2011a}, we consider $a$ as a positive constant to dominate the energy condition by the energy momentum tensor.\\

	The $V(r)$ of equation (\ref{metric_ansatz}) can be computed and takes the form as,
	\begin{equation}
		\label{metric_solution_V_with_K}
		V(r) = k + \frac{r^2}{l^2} -\frac{m}{r^{n-2}} + \frac{q^2}{r^{2n-4}}-\frac{16\pi G_{n+1} a}{(n-1)r^{n-3}}.
	\end{equation}
 Where $k= 1,0,-1$ depends on whether the $(n-1)$ dimensional boundary is spherical, flat or hyperbolic respectively. The $V(r)$ is written with the parametrisation of cosmological constant $\Lambda = -\frac{n(n-1)}{2 l^2}$, where $l$ corresponds to the radius of the AdS spacetime, $m$ is the constant of integration which represents the mass parameter of the black hole and is related to the ADM mass $(M)$ of the black hole. \\

\begin{figure}[h]
\begin{center}
\begin{psfrags}
\psfrag{V(r)}[][]{$ V(r)$}
\psfrag{r}[][]{$ r$}
\psfrag{a}[][]{$a$}
\psfrag{q}[][]{$q$}
\psfrag{m}[][]{$m$}
\psfrag{l}[][]{$l$}
\subfigure[]{\includegraphics[width=7.5cm]{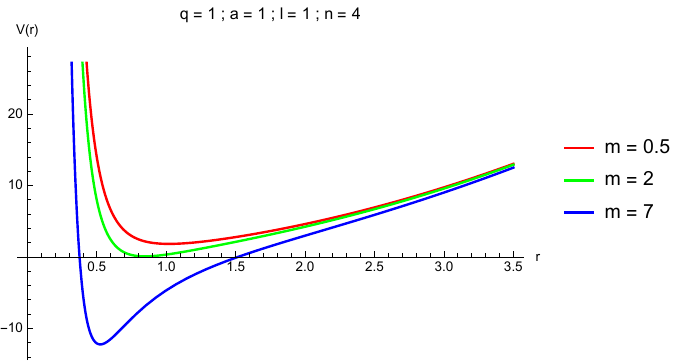}}
\quad
\subfigure[]{\includegraphics[width=7.5cm]{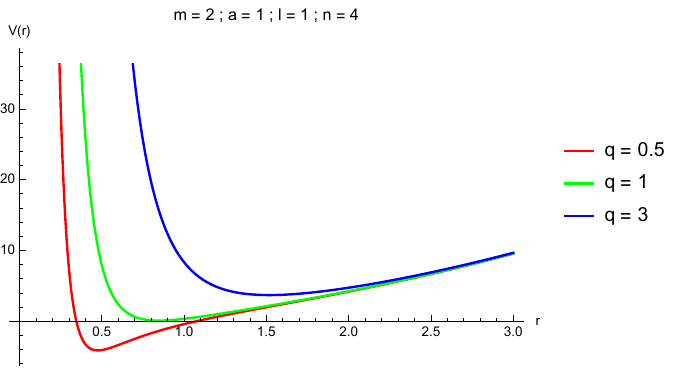}}
\end{psfrags}
\end{center}
\caption{ $V(r)$ vs $r$ plot, figure (a) for different values of $m$ and figure (b) for different values of $q$. All the graphs are drawn for $k = 1$.}
\label{horizon}
\end{figure}

The metric tensor corresponding to the $V(r)$ of equation (\ref{metric_solution_V_with_K}) has singularity at $r=0$ for the generic value of $n \ge 3$. Irrespective of the nature of spacetime curvature, from the Descartes' sign rules, it is clear that $V(r)=0$ equation have either two or no real positive roots just like a charged black hole.  Above the extremal condition there are two black hole horizons-- the inner and the outer horizon. The singularity of the metric tensor at $r=0$ is protected by these horizons. Below the extremal condition there is no horizon and singularity is called naked singularity. At extremal condition, the two horizons marge together and provide single horizon. The extremal condition depends on the value of parameters like $ m, q$ etc. as shown in the figure (\ref{horizon}).\\

The black hole mass parameter $ m $ can be calculated from the equation $V(r=r_+)=0$ and it can be written as,
	\begin{equation}
		\label{mass_m_equation}
		m = \frac{(n-1) r_+^{2n-2} + (n-1)l^2 k r_+^{2n-4}-16 \pi G_{n+1}a l^2 r_+^{n-1}+(n-1)l^2 q^2}{(n-1) l^2 r_+^{n-2}}.
	\end{equation}
Here $r_+$ is the horizon radius. The integration constant $m$ is related with the ADM mass $(M)$ of the black hole in the following way,
	 \begin{equation}
		 \label{ADM_mass_equation}
		 M = \frac{(n-1) \omega_{n-1}}{16 \pi G_{n+1}} m = \frac{(n-1) \omega_{n-1}}{16 \pi G_{n+1}}\left[\frac{r_+^n}{l^2} + k r_+^{n-2} + \frac{q^2}{r_+^{n-2}} -\frac{16 \pi G_{n+1}a r_+}{(n-1)}\right],
	 \end{equation}
	 where $\omega_{n-1}$ is the volume of the $(n-1)$ sphere. Due to the string cloud contribution, the mass of the black hole can be negative for a certain size of black hole even for $k = 1$.  Latter we show that the entropy of the black hole still maintain the one fourth of the area law. \\
	\section{Thermodynamical Quantities}
	\label{subsection_temperature}
	It is well-established that black holes can be treated as thermodynamic systems and therefore, we can calculate various thermodynamic quantities associated with them, such as temperature, entropy, and free energy. To begin, we will focus on calculating the temperature of the black hole.
\subsection{Temperature}
The temperature of the black hole is the inverse of the periodicity $\beta$ of Euclidean time $(\tau \rightarrow it)$ of the black hole metric. The temperature is given by,
	\begin{equation}
		\label{temperature_formula_for_black_holes}
		T = \frac{1}{4 \pi} \left.\frac{d V(r)}{dr}\right|_{r=r_+}.
	\end{equation}
Using (\ref{metric_solution_V_with_K}) it becomes,
	\begin{equation}
		\label{temperature_T_with_K_and_n}
		T = \frac{n(n-1)r_+^{2n+2}+ k l^2 (n-1)(n-2)r_+^{2n} - 16 \pi G_{n+1} l^2 a r_+^{n+3} - (n-1)(n-2) l^2 q^2 r_+^4}{4 \pi l^2 (n-1) r_+^{2n+1}}.
	\end{equation}
	The Descartes' sign rules indicate that there exists an extremal black hole solution with a radius, denoted as $r_{\rm vac}$, even at zero temperature and charge $q = 0$, regardless the value of $k$. The extremal radius $r_{vac}$ is the solution to the equation,
\begin{equation}
	\label{temperature_condition_cubic_equation}
n(n-1) r_{\rm vac}^n + (n-1) (n-2) k l^2 r_{\rm vac}^{n-2}  -16 \pi G_{n+1} a l^2 r_{\rm vac}=0.
\end{equation}
When the string density disappears, $r_{\rm vac}$ becomes zero and the background becomes the AdS spacetime. The mass parameter of the extremal black hole is given by
\begin{equation}
		\label{mass_m0_equation}
		m_0 = \frac{1}{n}\left[2 k r_{\rm vac}^{n-2}-16 \pi G_{n+1}a r_{\rm vac}\right].
	\end{equation}
The extremal black hole mass can also be negative for small value of $r_{\rm vac}$ even for $ k = 1$ like negative curvature spacetime \cite{Vanzo1997}.

In order to compute other thermodynamical quantities such as free energy, entropy, etc. one can use the first law of thermodynamics considering it valid in this configuration. However, we here evaluate the on shell Euclidean action and then show that the first law of thermodynamics still hold for this configuration and finally calculate the above quantities.


\subsection{Action Calculation}
\label{section_action_calculation}
In this section, we aim to calculate the on-shell Euclidean action for the system described by equation (\ref{action_total}). The on shell Euclidean action usually diverges for infinite volume of the spacetime. The action must be finite to get finite values of the thermodynamic quantities. Therefore, to have a finite value of the on shell Euclidean action, we can either add a counter term \cite{Balasubramanian1999} with the action or following Witten \cite{Witten2014}, subtract a contribution of the action of another spacetime which has the same asymptotic geometry. We opt for the latter method to compute the on shell action. In \cite{Witten2014}, Witten computed the on-shell Euclidean action of AdS Schwarzshild black hole by subtracting the action of the AdS spacetime. The following steps are taken in his computation. First set a limit on the radius of the spacetime up to a certain value $r= r_{\rm max}$ and subtract the contribution of the action of the AdS spacetime from the action of the AdS Schwarzshild black hole and then take the limit $r_{\rm max} \rightarrow \infty$.\\

 For charged AdS black hole, in \cite{Chamblin1999a}, the thermal AdS is chosen as the reference background to match the geometry of the AdS and black hole in the asymptotic region. The thermal AdS background is obtained by setting $m=q=0$ in the metric solution of the charged black hole.\\

Before going to compute the Euclidean action following \cite{Anninos2009}, we take the trace of the equation of motion (\ref{equation_of_motion_wrt_g_munu}) and rearrange the terms to match the structure of the action given in equation (\ref{action_total}) which will be useful later to calculate the action. The resulting expression for the trace of the equation of motion is,
\begin{equation}
  R-\frac{(n+1)}{2}R + (n+1)\Lambda = 2 F^{\nu\lambda}F_{\nu\lambda}-\frac{(n+1)}{2}F^2 + 8 \pi G_{n+1} T^{SC}.
\end{equation}
After rearranging the terms, we get the above equation in the following form,
\begin{equation}
\label{actionform}
R-2\Lambda -F^2 = -\frac{2}{(n-3)}[ R - 4 \Lambda + 8 \pi G_{n+1}  T^{SC}].
\end{equation}
Now, we compute the on shell Euclidean action in the following way,
\begin{equation}
\label{Eclideanaction}
I_E  = \frac{1}{8\pi  G_{n+1}(n-3)}\int d^{n+1}x \sqrt{-g} \left[ R - 4 \Lambda + 8 \pi G_{n+1} T^{SC}\right]+ \int d\tau dr a \omega_{n-1}.
\end{equation}
Here, the on shell string action is expressed as,
\begin{equation}
	I_{SC} = -\int d^{n+1}x \, a,
\end{equation}
since we have incorporated the delta function in the density function $a(x)$ in (\ref{string_density_equation}). The above equation (\ref{Eclideanaction}) can be recasted as,
\begin{equation}
I_E  =  \frac{\omega_{n-1}\beta}{8\pi G_{n+1}(n-3)}\int dr r^{n-1} \left[R - 4 \Lambda + 8 \pi G_{n+1} T^{SC}\right]+ \omega_{n-1}\beta\int dr a.
\end{equation}
After replacing the value of $T^{SC}$ by using equation (\ref{string_cloud_non_vanishing_components}), we can write the above equation as,
\begin{equation}
I_E  = \frac{\omega_{n-1}\beta}{8\pi G_{n+1}(n-3)}\int dr  \left[ R r^{n-1} - 4 \Lambda r^{n-1} +  8\pi  G_{n+1} (n-5)a \right].
\end{equation}
As mentioned in the beginning of this section, the value of the integral is infinite for the infinite volume or infinite radius $r$ of the spacetime. In order to avoid this infinity we take $r= r_{\rm max}$ and subtract the contribution of the background called vacuum $(\rm vac)$ whose boundary is similar to our spacetime boundary.  Due to the  contribution from the string cloud, irrespective of the value of $k$, metric solution has a horizon and corresponding temperature for $m=q=0$. In order to avoid the problem of the reference background having a fixed temperature and keeping in mind that the black hole mass is negative for small size,  we choose the neutral extremal black hole obtained by setting $T = q = 0$ which has also negative mass for small size. Finally, we take $r_{\rm max}\rightarrow \infty$. By doing this, we can write the Euclidean action as,
\begin{align}
\label{aftersubtraction}
I_E  = \frac{\omega_{n-1}}{8\pi G_{n+1}(n-3)}&\left(\beta\int_{r_+}^{r_{\rm max}} dr  \left[ R r^{n-1} - 4 \Lambda r^{n-1} +  8\pi G_{n+1}(n-5) a\right]\right. \nonumber \\  &- \left.\beta_{\rm vac}\int_{r_{\rm vac}}^{r_{\rm max}}dr  \left[R_0 r^{n-1} - 4 \Lambda r^{n-1} +  8\pi  G_{n+1}(n-5)a \right]\right).
\end{align}
Here $\beta_{\rm vac}$ is the Euclidean time periodicity of the vacuum and $R_0$ is the Ricci scalar for the background.  In order to have the same boundary between the black hole and vacuum spacetime, the following constraint must be satisfied,
\begin{equation}
\sqrt{V(r_{\rm max})}\beta = \sqrt{ V_{\rm vac}(r_{\rm max})}\beta_{\rm vac}.
\end{equation}
Here $V(r_{\rm max})$ and  $V_{\rm vac}(r_{\rm max})$ are the time component of the spacetime metric at the boundary of the black hole  and vacuum spacetime respectively. Finally, the Euclidean action of equation (\ref{aftersubtraction}) takes the form as,
\begin{align}
\label{actioninv}
I_E = \frac{\beta\omega_{n-1}}{8\pi G_{n+1}(n-3)}&\left(\int_{r_+}^{r_{\rm max}} dr  \left[R r^{n-1} - 4 \Lambda r^{n-1} +  8\pi G_{n+1}(n-5)a \right]\right. \nonumber \\  &- \left.\sqrt{\frac{V(r_{\rm max})}{V_{\rm vac}(r_{\rm max})}}\int_{r_{\rm vac}}^{r_{\rm max}}dr  \left[R_0 r^{n-1} - 4 \Lambda r^{n-1} +  8\pi  G_{n+1}(n-5) a \right]\right).
\end{align}
We define the $V_{\rm vac}(r)$ from $V(r)$ by setting the extremal condition $T = q = 0$. We also consider the form of Ricci scalar as,
\begin{equation}
R=-\frac{1}{r^{n-1}}(r^{n-1}V(r))'' + \frac{(n-1)(n-2)k}{r^2}.
\end{equation}
For background Ricci scalar $R_0$, we use $V_{\rm vac}(r)$ in place of $V(r)$ in the above equation. After substituting the value of $R$ in the equation (\ref{actioninv}), the action can be expressed as,
{\small
{
\begin{align}
I_E=\frac{\beta \omega_{n-1}}{8\pi G_{n+1} (n-3)} \left(\int_{r_+}^{r_{\rm max}}dr \left[-(r^{n-1} V(r))'' + (n-1)(n-2) k r^{n-3}-4\Lambda r^{n-1}+ 8\pi G_{n+1} (n-5) a\right]\right.\nonumber\\-\left(1-\frac{(m-m_0)l^2}{2 r_{\rm max}^n}\right)\left.\int_{r_{\rm vac}}^{r_{\rm max}}dr\left[-(r^{n-1} V_{\rm vac}(r))'' + (n-1)(n-2) k r^{n-3}-4\Lambda r^{n-1}+ 8\pi G_{n+1} (n-5) a\right] \right).
\end{align}
}}
After integration we get,
\begin{align}
\label{ndimaction}
I_E = \frac{\beta \omega_{n-1}}{8\pi G_{n+1} (n-3)}\left(\frac{(n-1)}{2}(m-m_0)+4\pi T r_+^{n-1}-(n-1)k r_+^{n-2} + \frac{4\Lambda }{n}r_+^n \right.\nonumber \\ -\left. 8 \pi G_{n+1} (n-5) a r_+   + (n-1)k r_{\rm vac}^{n-2} - \frac{4\Lambda }{n}r_{\rm vac}^n + 8 \pi G_{n+1} (n-5) a r_{\rm vac}\right).
\end{align}
Using the extremal condition of equation (\ref{temperature_condition_cubic_equation}), we can reduce the equation to,
\begin{align}{\label{finalaction}}
I_E &= \frac{\beta \omega_{n-1}}{16\pi G_{n+1} (n-3)}\left[(n-1)(m-m_0)+8\pi T r_+^{n-1}-2(n-1)k r_+^{n-2} + \frac{8\Lambda }{n}r_+^n \right.\nonumber \\ &- \left. 16 \pi G_{n+1} (n-5) a r_+   + (n-1)(n-3)(n-4)k r_{\rm vac}^{n-2} - \frac{2 (n-1)(n-4) \Lambda }{n}r_{\rm vac}^n\right].
\end{align}

Now we are ready to compute the thermodynamical quantities like entropy, specific heat, total internal energy and free energy etc.


\subsection{Other Thermodynamical Quantities}
In order to determine the state variables of the black hole spacetime we consider potential $\Phi$ is fixed at the boundary. Also following \cite{Kumar:2012ui}, we consider string density is unchanged since strings are infinitely heavy. Therefore there will not be any contribution from the string clouds in the free energy expression. Finally, the free energy is expressed as $ F=E-TS-Q\Phi$. Following \cite{Chamblin1999a}, we can calculate the state variables of the system from the Euclidean action of equation (\ref{finalaction}) as \footnote{The thermodynamic quantities have been calculated by expressing the horizon radius in terms of temperature, charge etc. For more details see \cite{Anninos2009,Chamblin1999a}.};\\

\noindent
\begin{itemize}
\item Entropy:
\begin{equation}
	\label{entropy_relation_and_value_from_action}
	S = \beta\left.\left(\frac{\partial I_E}{\partial \beta}\right)\right|_\Phi -I_E = \frac{r_+^{n-1} \omega_{n-1}}{4 G_{n+1}}.
\end{equation}
\item Charge:
\begin{equation}
	\label{charge_relation_and_value_from_action}
	Q = -\frac{1}{\beta}\left.\left(\frac{\partial I_E}{\partial \Phi}\right)\right|_\beta = \sqrt{2 (n-1)(n-2)}\left(\frac{\omega_{n-1}}{8 \pi G_{n+1}}\right)q.
\end{equation}
\item Energy:
\begin{equation}
	\label{energy_relation_and_value_from_action}
	E = \left.\left(\frac{\partial I_E}{\partial \beta}\right)\right|_\Phi-\frac{\Phi}{\beta}\left.\left(\frac{\partial I_E}{\partial \Phi}\right)\right|_\beta = M + E_0.
\end{equation}
\end{itemize}
Where $M$ is the ADM mass defined in (\ref{ADM_mass_equation}) and $E_0$ is the background energy contribution,
\begin{equation}
\label{background_energy_E_o}
	E_0 =\frac{(n-1)\omega_{n-1}r_{\rm vac} }{16 \pi G_{n+1}n(n-3)l^2}\left[n(n-1)(n-4)r_{\rm vac}^{n-1} + kl^2 \{n(n-3)(n-4)-2\}r_{\rm vac}^{n-3} + 16\pi G_{n+1} a l^2 \right].
\end{equation}
For $r_{\rm vac} = 0$ spacetime, $E_0$ becomes zero and for $n = 4 $ spatial dimensions it turns out to be
\begin{equation}\label{4dimezero}
  E_0 = -\frac{3\, \omega_3}{32 \pi G_5}\left[ k r_{\rm vac}^2 - 8 \pi G_5 a r_{\rm vac}\right] = -\frac{3\, \omega_3}{16 \pi G_5} m_0.
\end{equation}
For the purpose of latter convenience we write $E_0$ in terms of only $r_{\rm vac}$ using the extremal condition of equation (\ref{temperature_condition_cubic_equation}) and it takes the form,
\begin{equation}\label{rvacezero}
  E_0 = \frac{(n-1)\omega_{n-1}r_{\rm vac}^{n-2}}{16 \pi G_{n+1}l^2}\left[ (n-3)kl^2 + (n-1)r_{\rm vac}^2 \right].
\end{equation}
These state variables naturally satisfy the first law of thermodynamics $ dE = TdS + \Phi dQ$. Therefore one can compute the specific heat of the black hole by using the relation,
\begin{align}
\label{specific_heat_formula}
C &= T\left(\frac{\partial S}{\partial T}\right)\nonumber\\& = \frac{(n-1)\omega_{n-1} r_+^{n-1}}{4 G_{n+1}}\frac{[n(n-1) r_+^{n-1} + l^2 [k(n-1)(n-2)-2(n-2)^2 \Phi^2]r_+^{n-3} - 16\pi G_{n+1} l^2 a]}{[n(n-1)r_+^{n-1} - l^2 [k (n-1)(n-2) - 2 (n-2)^2 \Phi^2]r_+^{n-3} + 16\pi G_{n+1} l^2 (n-2) a]}.
\end{align}
\begin{itemize}
\item Free energy:

We compute the free energy of the black hole from the thermal partition function Z
of the black hole solution. The thermal partition function can be calculated from the classical action of the Euclidean black hole metric solution by using the saddle point approximation through the following relation,
\begin{equation}
		\label{partionfun}
 Z = e^{-I_E}.
\end{equation}
Therefore, the free energy of the black hole can be written as,
\begin{equation}
\label{partionfun1}
F = -k_B T\, {\rm ln} Z = k_B T I_E,
\end{equation}
where $I_E$ is the Euclidean action of the black hole solution. If Boltzmann constant $k_B $ is considered to be equal to one and $T = \frac{1}{\beta}$, then equation (\ref{partionfun1}) reduces as,
\begin{equation}
		\label{partionfun}
F \beta =   I_E.
\end{equation}
Therefore the free energy of the black hole takes the form as,
\begin{align}
		\label{free energy equation}
		F &= \frac{I_E}{\beta}  = E-T S- \Phi Q\nonumber\\ & = \frac{\omega_{n-1}}{48 \pi G_{n+1}}\left(-\frac{3}{l^2} r_+^n + \frac{3[k(n-1)- 2 (n-2) \Phi^2]}{(n-1)}r_+^{n-2} -\frac{48 \pi G_{n+1}(n-2) a }{(n-1)}r_+ \right) + E_0.
	\end{align}

\item Landau Function:
\end{itemize}
	
Now to calculate the Landau function, let us make the ansatz in terms of the order parameter $r_+$, going from the highest to the lowest power of the order parameter as given below,
	\begin{equation}
		\label{Landau function ansatz}
		G(r_+,T)= \frac{\omega_{n-1}}{48 \pi G_{n+1}}(b_n r_+^n - b_{n-1} T r_+^{n-1} + b_{n-2} r_+^{n-2} - b_{n-3} r_+^{n-3} + \cdots + (-1)^{n-1} b_1 r_+) + b_0.
	\end{equation}
	Here $b_n,b_{n-1},b_{n-2},b_{n-3}, \cdots , b_0$ are constants and can be fixed in the following way,
\begin{enumerate}
\item By substituting the temperature expression of equation (\ref{temperature_T_with_K_and_n}) in the Landau function expression of equation (\ref{Landau function ansatz}), we should get the free energy expression of equation (\ref{free energy equation}). Therefore, the existing term in the Landau function can be expressed as,
   \begin{align}
		\label{Landau function for the comparision with free enrgy}
		G(r_+,T)=\frac{\omega_{n-1}}{48 \pi G_{n+1} }&\left[\left( b_n - \frac{n b_{n-1}}{4 \pi l^2} \right) r_+^n \right. \nonumber\\&+ \left.\left(\frac{(n-2) [k(n-1) -2 (n-2) \Phi^2] b_{n-1}}{4\pi (n-1)} \right)r_+^{n-2}\right.\nonumber\\&+\left.\left(  \frac{4 G_{n+1} b_1 a }{(n-1)}+ (-1)^{n-1} b_1\right)r_+\right]+b_0.
	\end{align}

\item On the other hand, we take the derivative of the Landau function expression of equation (\ref{Landau function ansatz}) with respect to horizon radius $r_+$ and equate it with zero.
\begin{equation}
		\frac{\partial G}{\partial r_+}=0.
	\end{equation}
\item The solution of this equation gives the expression of temperature in terms of horizon radius $r_+$. Again the expected temperature expression can be written as,
\begin{equation}
		\label{T from landau function}
		T=\frac{n b_n}{(n-1) b_{n-1}} r_+ + \frac{(n-2) b_{n-2}}{(n-1) b_{n-1}}\frac{1}{r_+}+\frac{(-1)^{n-1} b_1}{(n-1) b_{n-1}}\frac{1}{r_+^{n-2}}.
	\end{equation}

\item Now, comparing the temperature expression of equation (\ref{T from landau function}) with the temperature expression of equation (\ref{temperature_T_with_K_and_n}) and Landau function expression of equation (\ref{Landau function for the comparision with free enrgy}) with the free energy expression of equation (\ref{free energy equation}), we get the non vanishing constants as,
\end{enumerate}
\begin{equation}
		\label{coefficient a1 value}
		b_1=-\frac{48\pi G_{n+1} a}{(-1)^{n-1}},
	\end{equation}
	\begin{equation}
		\label{coefficient a2 value}
		b_{n-2}= 3 [k(n-1) - 2 (n-2) \Phi^2],
	\end{equation}
	\begin{equation}
		\label{coefficient a3 value}
		b_{n-1}= 12 \pi,
	\end{equation}
	\begin{equation}
		\label{coefficient a4 value}
		b_n=\frac{3(n-1)}{l^2},
	\end{equation}
\begin{equation}
		\label{coefficient a4 value}
		b_0=E_0.
	\end{equation}
	Now using the values of non vanishing constant in (\ref{Landau function ansatz}) we get the Landau function as,
	\begin{align}
		\label{Landau function final}
		G(r_+,T)= \frac{\omega_{n-1}}{48\pi G_{n+1} }\left[ \frac{3(n-1)}{l^2} r_+^n - 12 \pi T r_+^{n-1} + 3 \{k(n-1) - 2 (n-2) \Phi^2\} r_+^{n-2}\right. \nonumber\\ - \Bigl.48 \pi G_{n+1} a r_+  \Bigr] + E_0.
	\end{align}
In the next section, we study the thermodynamic stability of the black hole with the help of the above computed thermodynamical quantities.


\section{Black Hole Phases}
\label{subsection_black_hole_phases}
\begin{figure}[h]
 \begin{center}
\begin{psfrags}
\psfrag{T}[][]{$\pi T$}
\psfrag{\phi}[][]{$ \Phi$}
\psfrag{a}[][]{$\bar a$}
\includegraphics[width=9cm]{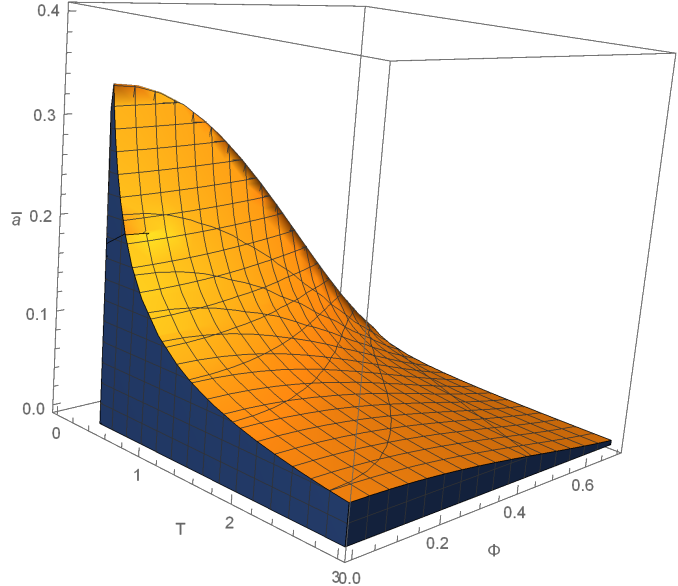}
\end{psfrags}
\end{center}
\caption{Parameter space of $T$, $\Phi$ and $\bar{a}$ for positive discriminant.}
\label{discr}
\end{figure}

In this section, we study the different phases of black holes by analysing the thermodynamical quantities derived in the previous section. First, we scale the parameters as,
\begin{equation}
8 \pi G_5 =1,\,	\bar{r} = \frac{r_+}{l},\,\bar{a} = \frac{a}{l^2},\, \bar{q}=\frac{q}{l^2}.
\end{equation}
With these scaled parameters and for $ k = 1$, in five dimensions the Hawking temperature in equation (\ref{temperature_T_with_K_and_n}) of the black hole can be expressed as,
\begin{equation}\label{tempbar}
	T = \frac{(3 \bar r +  6 \bar{r}^3 -4 \bar r \Phi^2 -\bar a l) }{6 \pi l \bar{r}^2}.
\end{equation}
Now we plot the temperature $T$ with respect to the black hole radius $\bar r$ to get the information about the number of black hole solutions in a particular temperature.
\begin{figure}[h]
\begin{center}
\begin{psfrags}
\psfrag{T}[][]{$ T$}
\psfrag{r}[][]{$\bar r$}
\psfrag{a}[][]{$\bar a$}
\psfrag{\Phi}[][]{$ \Phi$}
	\mbox{
	\subfigure[]{\includegraphics[width=7.0cm]{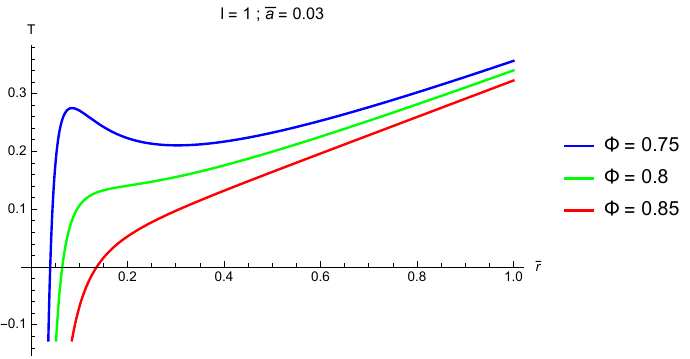}}
\quad
	\subfigure[]{\includegraphics[width=7.0cm]{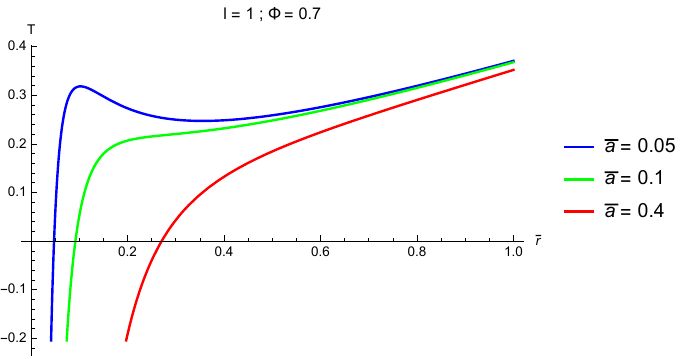}}}
\end{psfrags}
\end{center}
\caption{Both the graphs are Temperature ($ T $) vs. scaled horizon radius ($\bar r$). In figure (a) $\bar a$ is constant but in figure (b) $\Phi$ is constant.}
\label{tempar}
\end{figure}
Figure (\ref{tempar}a) illustrates that if the potential is lower than a certain critical value, there exists only one type of black hole solution upto a certain temperature, referred to as the small black hole. This solution persists even at zero temperature, known as the extremal black hole. At higher temperatures, two additional black hole solutions nucleate, with one referred to as the medium and the other as the large black hole, depending on their size. These three types of black holes coexist within a specific temperature range until, at a particular temperature, the small and medium black holes merge and disappear, leaving only the large black hole solution at high temperatures. If the potential exceeds the critical value, there is always one black hole solution present for any temperature. Figure (\ref{tempar}b)  exhibits similar behaviour but for different values of the scaled string cloud density $ \bar a$. \\
In order to know the range of potential $\Phi$ and scaled string density $ \bar a$ for which three black hole solutions exist together, we compute the discriminant of the cubic equation of the scaled horizon radius $\bar r$ from equation (\ref{tempbar}) which turns out to be,
\begin{equation}
	\label{discriminant_equation}
	\Delta = -972 \bar{a}^2 - 24 (3-4 \Phi^2)^3 + 648 \bar{a} (3-4 \Phi^2)(T\pi) + 36 (3-4 \Phi^2)^2 (T\pi)^2 - 864 \bar{a} (T \pi)^3.
\end{equation}
When the discriminant is greater than zero, it indicates the existence of three black hole solutions. Otherwise, there is only one black hole solution. In figure (\ref{discr}), the parameter space for temperature, potential, and string density is depicted, in which the discriminant becomes positive. The figure highlights that for a specific value of $\bar{a}$ and $\Phi$, there are two values of temperature for which the discriminant is positive. These temperatures set the upper and lower boundaries of the temperature range where three black hole solutions exist. If the potential $\Phi$ increases, the critical value of $\bar{a}$ decreases and becomes zero when $\Phi$ equals to the critical value of $\frac{\sqrt 3}{2}$ (for details refer to \cite{Chamblin1999a}). On the other hand when $\Phi$ takes zero value then $\bar a$ reaches its maximum value $\frac{1}{\sqrt 6}$ (for details see \cite{Dey:2020yzl, Dey2018}).\\

\begin{figure}[h]
 \begin{center}
\begin{psfrags}
\psfrag{T}[][]{$ T$}
\psfrag{C}[][]{$ C$}
\psfrag{a}[][]{$\bar a$}
\psfrag{P}[][]{$ \Phi$}
	\mbox{
\subfigure[]{\includegraphics[width=7.0cm]{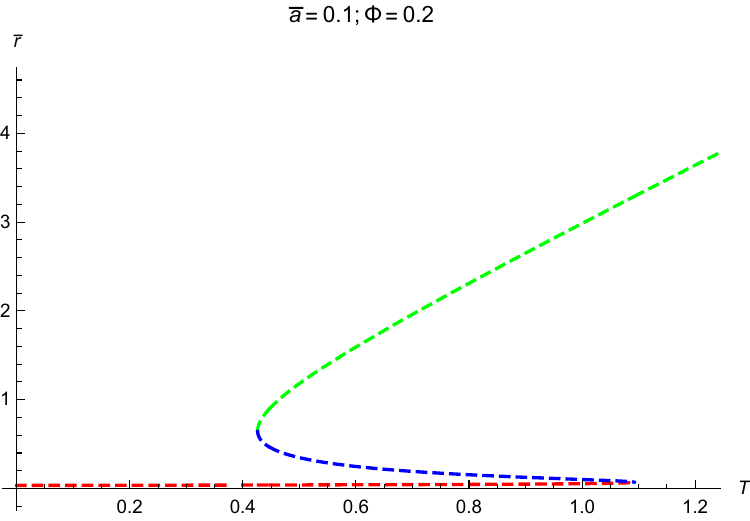}}
\quad
	\subfigure[]{\includegraphics[width=7.0cm]{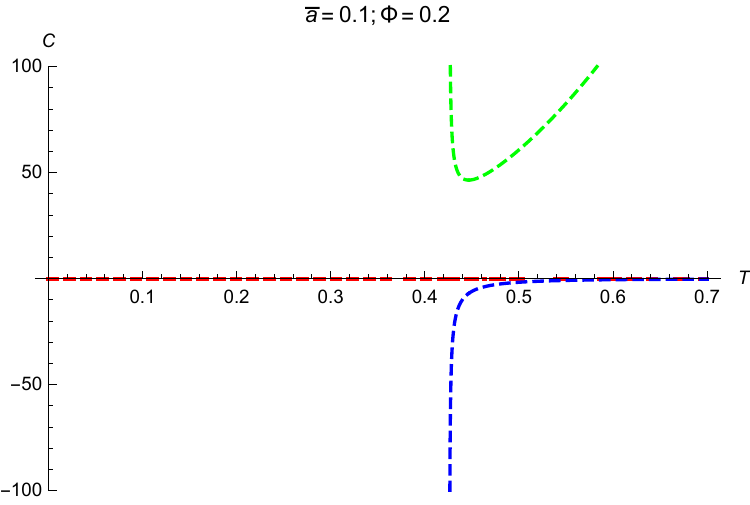}}}
\end{psfrags}
\end{center}
\caption{Figure $(a)$ and $(b)$ are drawn for the value of $\Phi$ and $\bar a$ within their critical values. Figure $(a)$ is for number of solution with respect to temperature $T$ and figure $(b)$ for the specific heat with respect to temperature $T$. }
\label{specificheat}
\end{figure}

Having conducted an analysis on the existence of various black hole solutions in a specific parameter space, our next objective is to evaluate their stability. To accomplish this, we focus on the three black hole solutions present in the parameter space and investigate their specific heat, free energy, and Landau function. We begin our analysis by calculating the specific heat using the formula,
\begin{equation}
C =  \frac{6 \pi \omega_3 l^3 \bar{r}^3 \big[ 6 \bar{r}^3 + 3 \bar r (1-\frac{4}{3} \Phi^2)-l\bar{a}\big]}{\big[6 \bar{r}^3 -3 \bar r (1-\frac{4}{3} \Phi^2) +2 l\bar a\big]}.
\end{equation}
Figure (\ref{specificheat}a) illustrates the relationship between the temperature and the number of black hole solutions. It shows that there are three black hole solutions within a specific temperature range, whereas outside of this range, only one black hole solution exists. In figure (\ref{specificheat}b), we study the specific heat with respect to the temperature.  We observe that the specific heat is positive for small and large black holes, but negative for medium-sized black hole. Consequently, we conclude that the medium-sized black hole is unstable, whereas the small and large black holes are stable.
\\

\begin{figure}[h]
\begin{center}
\begin{psfrags}
\psfrag{T}[][]{$T$}
\psfrag{F}[][]{$ F$}
\psfrag{a}[][]{$\bar a$}
\psfrag{r}[][]{$\bar r$}
\psfrag{P}[][]{$\Phi$}
	\mbox{
\subfigure[]{\includegraphics[width=7.5cm]{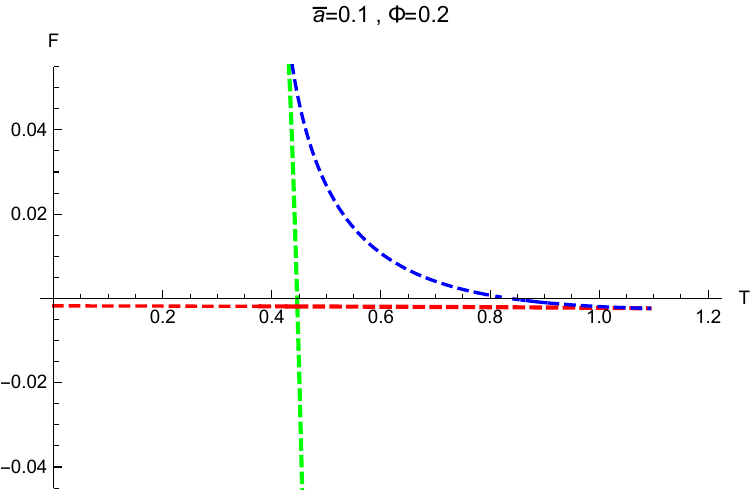}}
\quad
	\subfigure[]{\includegraphics[width=7.5cm]{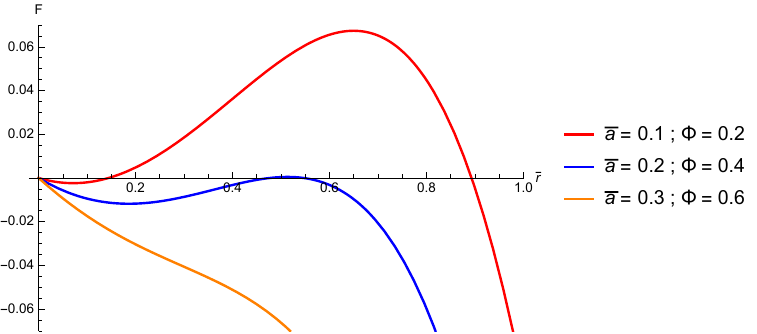}}}
\end{psfrags}
\end{center}
\caption{Figure $(a)$ is for free energy with respect to temperature $T$ and figure $(b)$ for the free energy with respect to scaled horizon radius $\bar{r}$. Both the graphs drawn within the critical values of $\Phi$ and $\bar a$.}
\label{freeenergy}
\end{figure}

Now we proceed to study the free energy with respect to the temperature and radius of the black hole. The free energy can be calculated from the action in equation (\ref{free energy equation}) and with the shift of ground state energy by $E_0$, the free energy takes the form as,
\begin{equation}
F=\frac{ \omega_3 l^2 \bar r}{2}\big[\bar r(1-\frac{4}{3}\Phi^2) - \bar r^3 -\frac{4}{3}l\bar a\big].
\end{equation}
 In figure (\ref{freeenergy}a) we present a plot of the free energy against temperature for a fixed value of scaled string cloud density $\bar a$ and potential $\Phi$.  At low temperatures, there is only one black hole solution, namely the small black hole, and it has negative free energy, denoted by the red dashed line. This indicates its stability at this low temperature. As the temperature increases, two more black hole solutions appear, corresponding to the medium and large black holes along with the small one. Initially, both the medium and large black holes possess positive free energy, with the green dashed line representing the large black hole and the blue dashed line representing the medium black hole. As the temperature increases, the free energy of both the black holes reduces, and eventually, at a critical temperature $ T_c$, the free energy of the large black hole becomes equivalent to the small black hole and goes lower at higher temperatures. Furthermore, the free energy of the medium black hole merges with the small black hole at another temperature, which is higher than $T_c$. In figure (\ref{freeenergy}b), we plot the free energy with respect to the scaled black hole horizon for different $ \bar a$ and $\Phi$.
\\

\begin{figure}[h]
 \begin{center}
\begin{psfrags}
\psfrag{T}[][]{$\pi T$}
\psfrag{P}[][]{$ \Phi$}
\psfrag{a}[][]{$\bar a$}
\includegraphics[width=8.5cm]{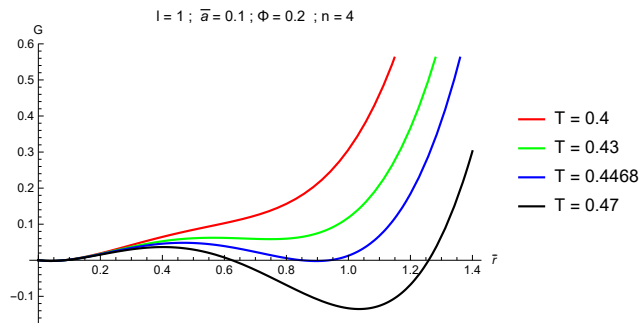}
\end{psfrags}
\end{center}
\caption{Landau function($G$) vs $\bar r$ plot for fixed $ \bar a$, $ \Phi$ and  different values of $T$.}
\label{Landauplot}
\end{figure}

\begin{figure}[h]
 \begin{center}
\begin{psfrags}
\psfrag{T}[][]{$\pi T$}
\psfrag{P}[][]{$ \Phi$}
\psfrag{a}[][]{$\bar a$}
\mbox{\subfigure[]{\includegraphics[width=7.0cm]{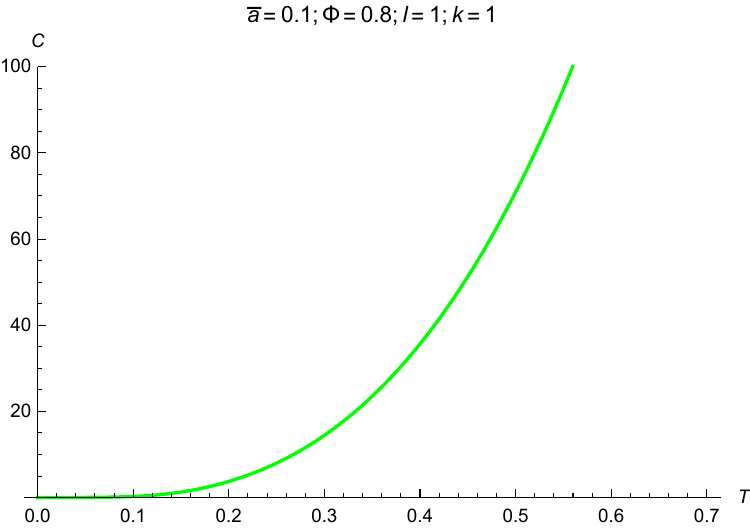}}
\quad
\subfigure[]{\includegraphics[width=7.5cm]{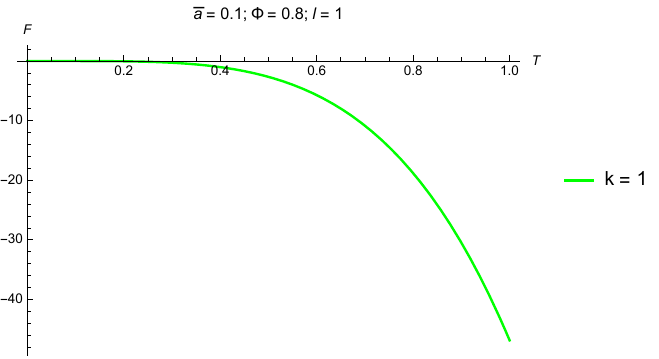}}}
\end{psfrags}
\end{center}
\caption{Figure $(a)$ is drawn for the specific heat with respect to temperature $T$ and figure $(b)$  is drawn for the free energy against temperature. Both the graph are drawn above the critical values of $\Phi$ and $\bar a$ }
\label{singlesol}
\end{figure}

Now we study the Landau function which can be expressed as,
\begin{equation}
G= \frac{3 \omega_3 l^2}{2}\left[\bar r^4 -\frac{4 \pi l T \bar r^3}{3}+ \bar r^2\left(1-\frac{4\Phi^2}{3}\right)-\frac{2 l \bar a \bar r}{3}\right].
\end{equation}
In figure (\ref{Landauplot}), we plot the Landau function with respect to scaled horizon $\bar r$ for different temperatures of the black hole. At low temperatures, only one black hole solution corresponding to the small black hole exists, while at higher temperatures, two more black holes nucleate, corresponding to the medium and large black holes, respectively, with a higher value of the Landau function. However, as the temperature continues to increase, the value of the Landau function decreases, and finally, above the critical temperature $T_c$, the Landau function of the large black hole becomes lower than that of the small black hole. Therefore, at low temperatures, the small black hole is stable, and at high temperatures, the large black hole becomes stable, with a critical temperature $T_c$ marking the transition between the small and large black holes. This is similar to the Hawking-Page like transition.\\
\\
Now we consider the  chemical potential $ \Phi$  and scaled string density $\bar a$ above their critical values so that only one black hole solution exist.
In figure (\ref{singlesol}a), we plot the specific heat  with respect to temperature. It shows that the specific heat is always positive. Therefore, the single black hole is stable configuration against AdS spacetime. In figure (\ref{singlesol}b), free energy is plotted with respect to temperature $T$, which shows that it is a stable configuration.
\\
Though we are not providing the data here for the flat and hyperbolic nature of the curvature of spacetime, we can similarly conclude that there exists only one stable black hole solution and further analysis is not necessary.\\

In absence of string cloud, the configuration reduces to the charged AdS black hole solution and there are unstable and stable black holes, those correspond to the medium and large black hole of our system. The phase transition occurs between thermal AdS and AdS black hole. However in our case due to the presence of long heavy strings, the AdS space is deformed and gives the presence of an extra small stable black hole along with medium and large black holes. Therefore, the phase transition occurs between small and large black hole instead of thermal AdS and AdS black hole.\\

In the overall analysis, AdS space is no longer a valid state in this configuration. Consequently, one can expect that in the corresponding dual gauge theory there might not be any confined state of quark-antiquarks. One should therefore study the dual theory side. As a preliminary step we examine the quark-antiquark distance against the position of the black hole horizon and Cardy-Verlinde formula in the following sections.


\section{Quark-antiquark distance}
\label{section_quark_antiquark_distance}

\begin{figure}[h]
\begin{center}
\includegraphics[width=8.0cm]{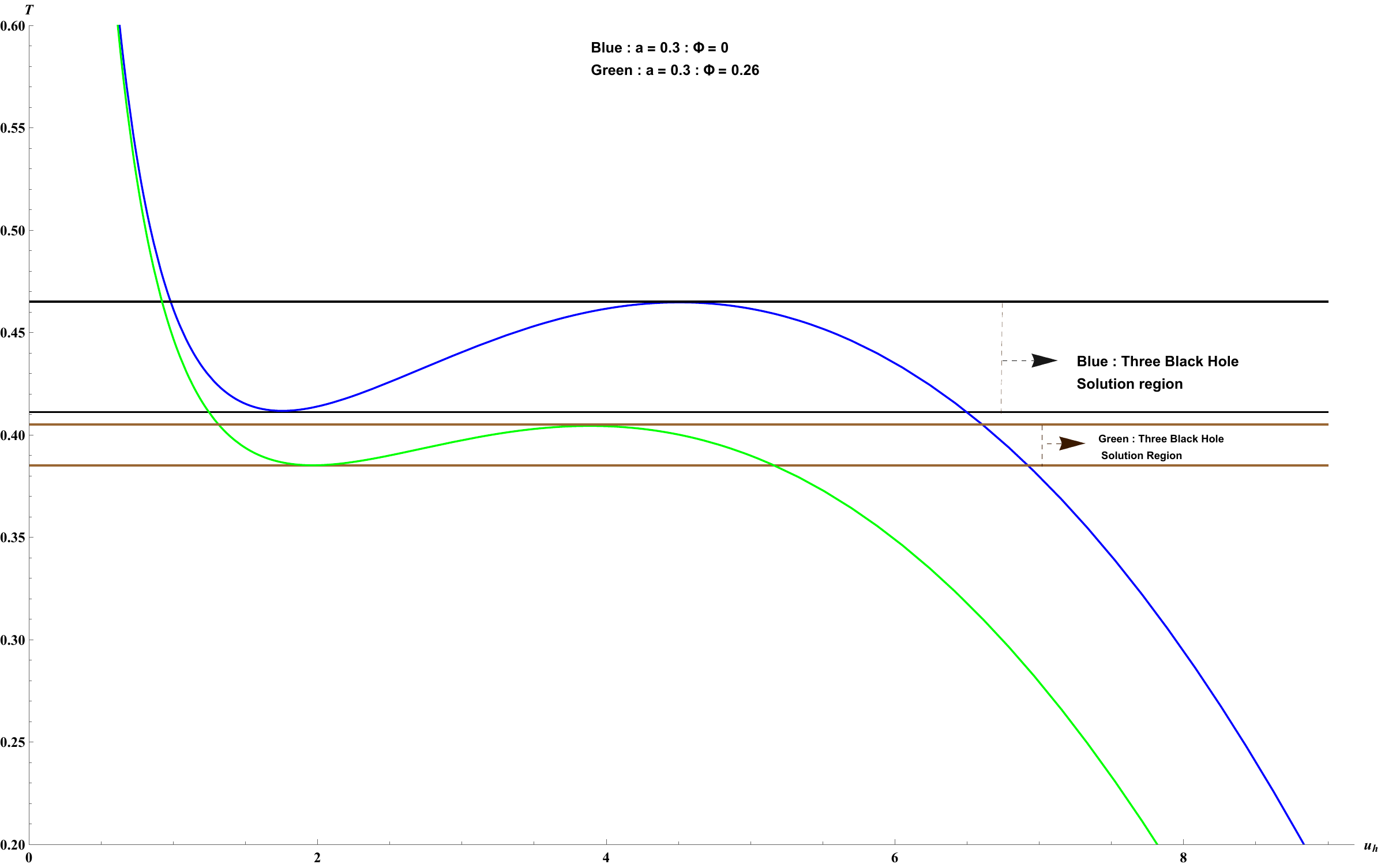}
\end{center}
\caption{Temperature $T$ vs $u_h$, Blue: $a = 0.3,\, \Phi = 0$ and Green : $ a = 0.3,\, \Phi =0.26$, corresponding three black holes solution region for both curves have been marked.}
	\label{Temp_Vs_uh_showing_region_of_BH_solution}
\end{figure}

It is interesting to examine the quark-antiquark$(q\bar q)$ distance in the boundary against the size of the black hole in the bulk theory as this could led to the understanding of the bound-unbound state of quark-antiquark pair in the dual theory. To study the $q\bar q$ distance, we consider a string configuration having its endpoints on the boundary and elongated in the bulk \cite{Dey:2020yzl,Dey2018}. Two kinds of  configurations of these strings are possible in bulk. A straight string in which the end points of the strings are attached on the boundary and hanging  up to the horizon or a U-shaped string where
	the string's end points are attached on the boundary and is hanging in the bulk with its tip at $u_0$ located in between boundary and horizon of the black hole in the bulk. Distance between the two endpoints of the string is denoted by $L$, which is the quark-antiquark distance.
	Quark-antiquark distance for a bound state can be obtained by considering a probe
	string in the black hole spacetime. The Nambu-Goto world sheet action for an open
	string is
	\begin{equation}
		S_{NG} = -\frac{1}{2 \pi \alpha^{'}}\int d^2 \sigma \sqrt{-det(h_{\beta \gamma})},
	\end{equation}
	where $\frac{1}{2 \pi \alpha^{'}}$
	is the string tension. The string tension is related to the ‘t Hooft coupling $\lambda$,
	in the dual SYM gauge theory as
	$$\sqrt{\lambda} = \frac{l^2}{\alpha^{'}}.$$
	We will assume that the above Nambu-Goto action is a valid action for our problem and
	corrections ensuing from coupling to the further bulk fields will not qualitatively modify
	the essential results.\\
	We will use $u$ rather than $r$, where $r =\frac{l^2}{u}$ so that $u = 0$ is the
	boundary and the horizon occurs at $u_h$. The metric tensor (\ref{metric_ansatz}) of the black hole solution for $n=4,\, k=1$ and $8\pi G_5=1$ reduces to the following form,
\begin{equation}
	ds^2 = f(u)(-h(u)dt^2 + dx^2 +dy^2 +dz^2 +\frac{du^2}{h(u)}),
\end{equation}
	where,
	\begin{align}
		f(u) &= \frac{l^2}{u^2},\\
		h(u) &= \left(1 +  \frac{u^2}{l^2}-\frac{m u^4}{l^6}+\frac{q^2 u^6}{l^{10}}-\frac{2 a u^3}{3 l^4}\right).
	\end{align}
The solution of the equation  $ h(u) = 0 $ gives the position of horizon $u_h$ of the black hole. The chemical potential at the black hole horizon takes the form as,
\begin{equation}
\Phi = \frac{{\sqrt 3} q u_h^2}{2l^4}
\end{equation}
and the temperature of the black hole is replaced as,
\begin{equation}
T = \frac{1}{6 \pi l^4 u_h}\Big[6 l^4 + 3 l^2 u_h^2 - 4 l^2 \Phi^2 u_h^2 - a u_h^3\Big].
\end{equation}
We plot the temperature with respect to the horizon radius of the black hole in the figure (\ref{Temp_Vs_uh_showing_region_of_BH_solution}). Here blue and green curves are drawn for zero and nonzero chemical potential respectively. The temperature ranges for small, large and three black hole solutions are marked for both the cases.

\begin{figure}[h]
	\centering
	\subfigure[]{\includegraphics[scale=0.2]{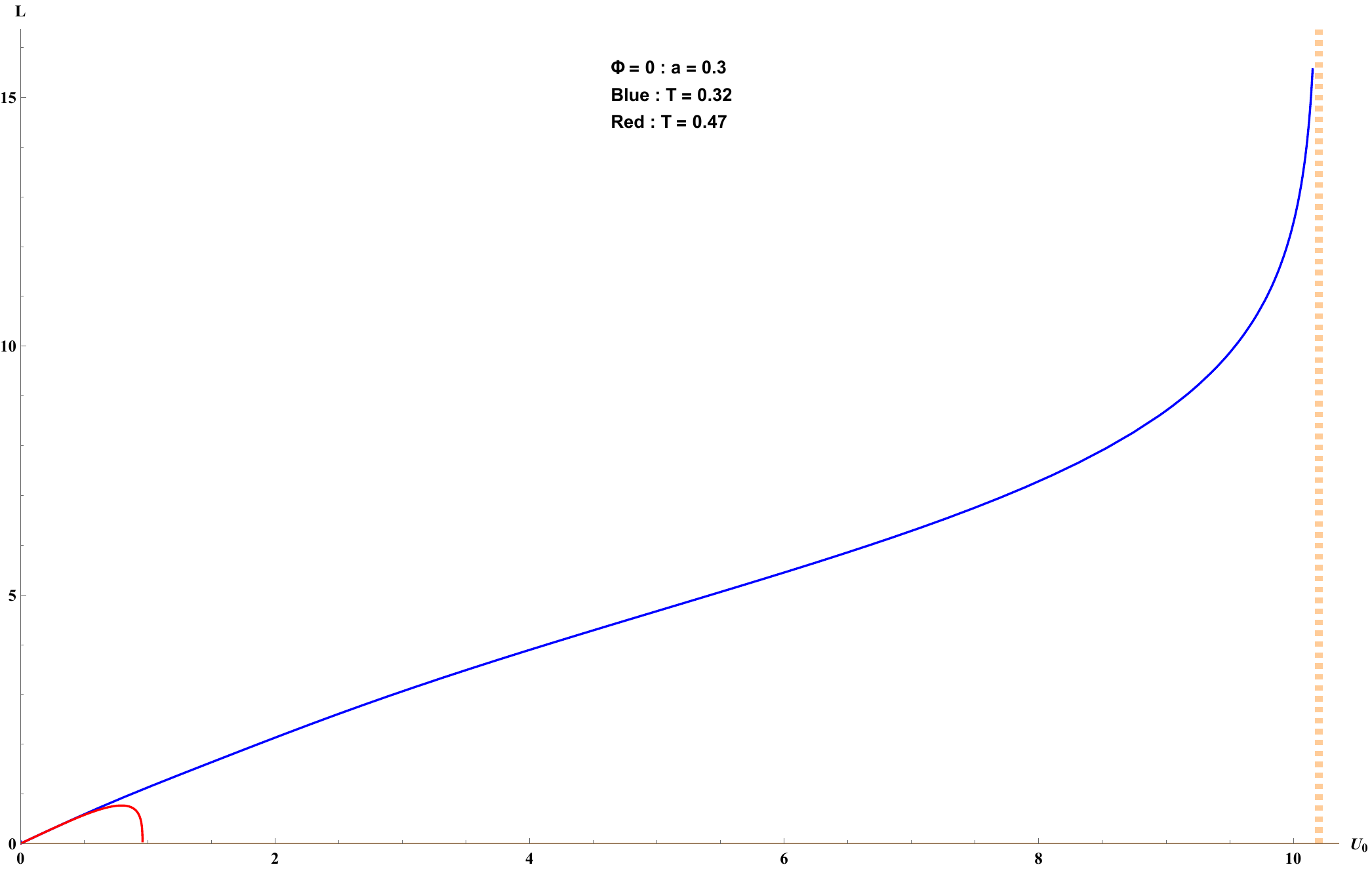}}
	\subfigure[]{\includegraphics[scale=0.2]{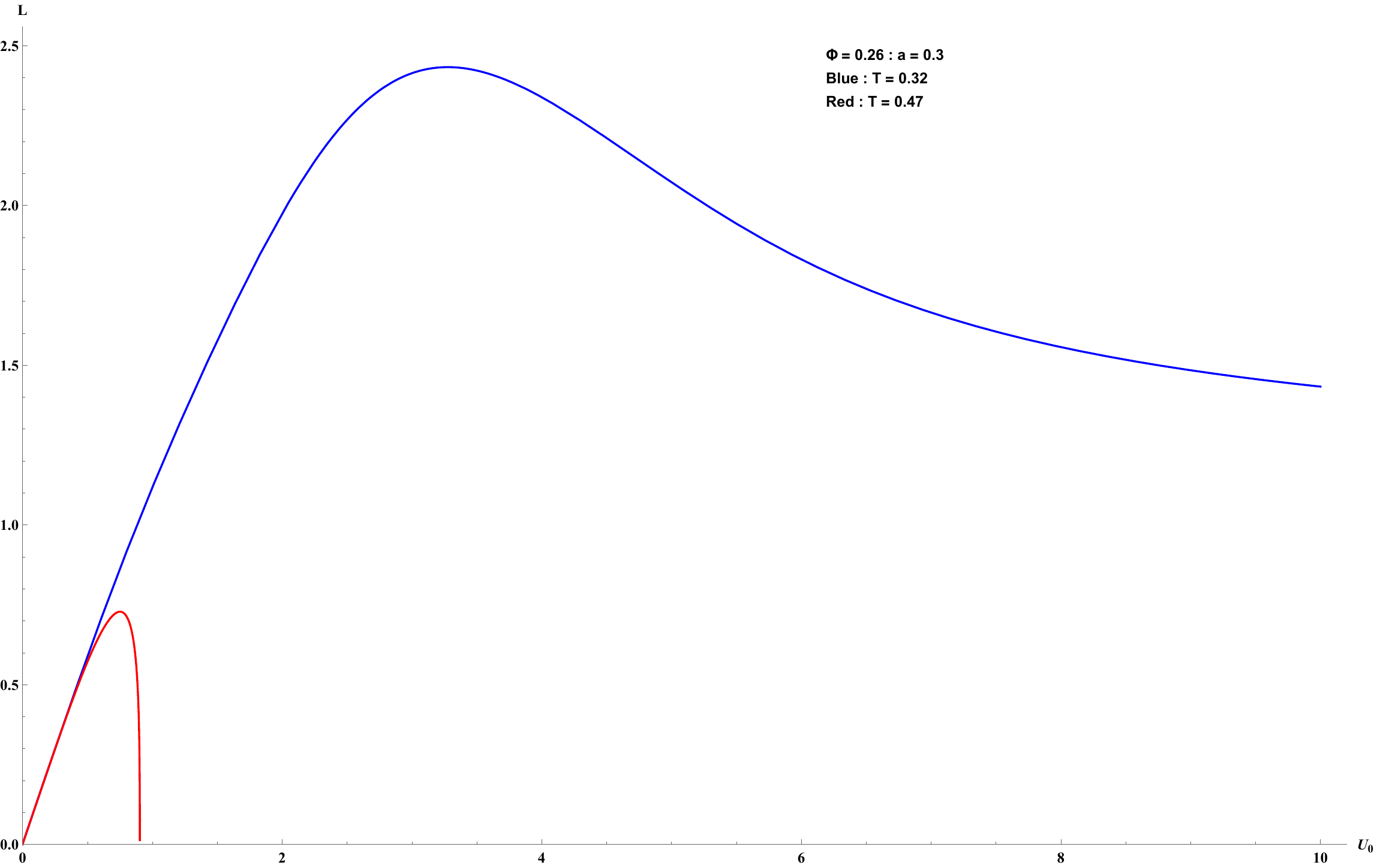}}
	\caption{Quark-antiquark distance $L$ Vs $u_0$: (a) $\Phi =0,\, a= 0.3$, $T=0.32$(blue) and $T=0.47$(red). (b) $\Phi =0.26,\, a= 0.3$, $T=0.32$(blue) and $T=0.47$(red). }
	\label{QAQ_dis_Vs_u0}
\end{figure}

\begin{figure}[h]
\begin{center}
	\includegraphics[width=8.0cm]{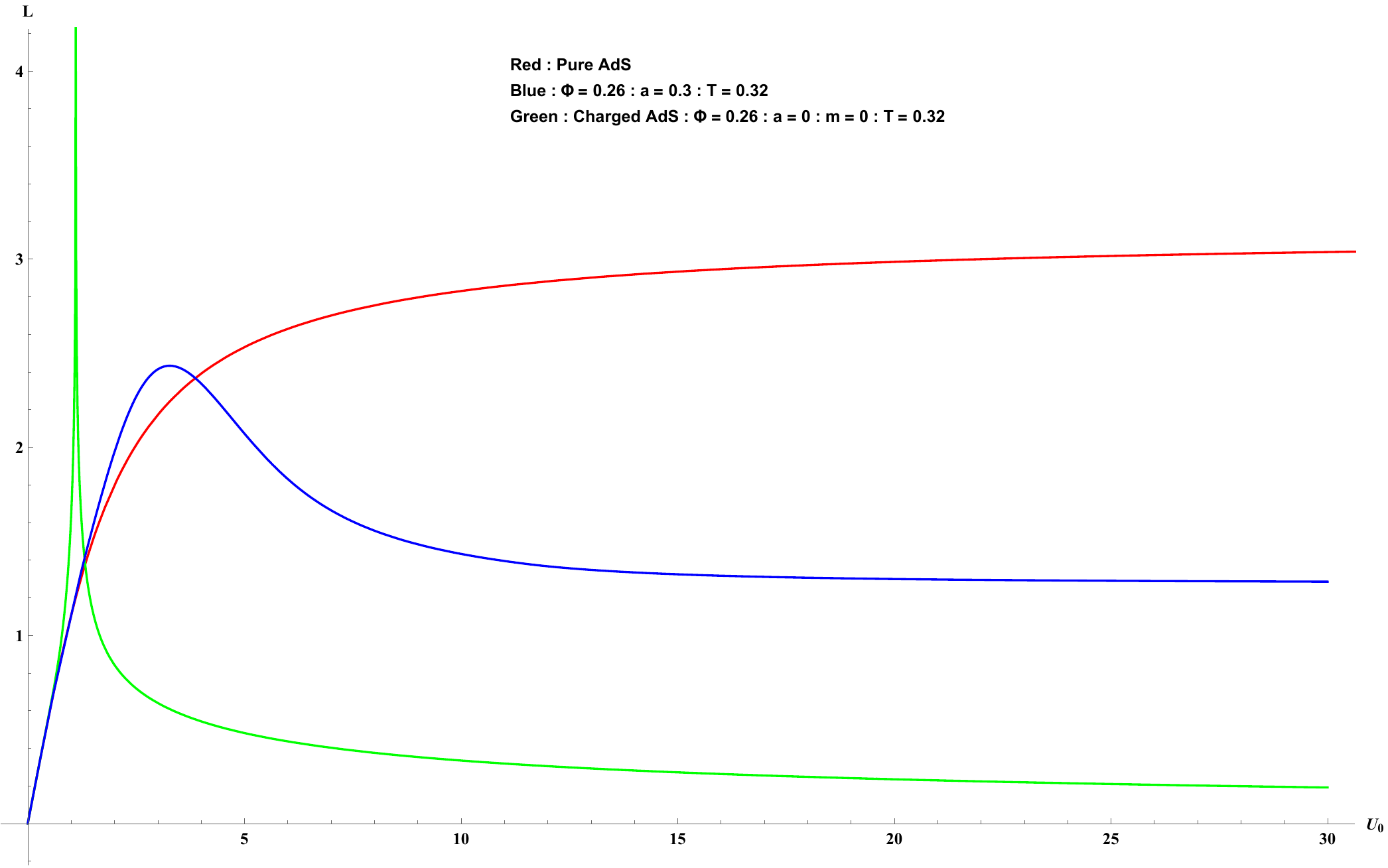}
\end{center}
	\caption{Quark-antiquark separation $L$ vs $u_0$, for $T = 0.32$, $a=0.3$ and $\Phi = 0.26$(blue), pure AdS solution (red) and charged AdS (green).}
	\label{QAQ_dis_Pure_AdS_and_nonzero_phi}
\end{figure}

The induced metric is $h_{\alpha \beta} = \partial_\alpha X^\mu \partial_\beta X^\nu g_{\mu \nu}$. We choose the following static gauge for the dual theory $\sigma^0 = t,\, \sigma^1 = x$, for these choices the induced metric in string frame can be written as
	\begin{equation}
		ds^2 = f(u)\left(-h(u) + (1 + \frac{u^{'2}}{h(u)})dx^2\right),
	\end{equation}
	where $u^{'}$ denotes a derivative w.r.t. $x$ .
	The Lagrangian and Hamiltonian of the quark and antiquark pair is obtained as,
	\begin{align}
		\mathcal{L} &= \sqrt{-det h_{\alpha \beta}} = f(u)\sqrt{h(u) + u^{'2}}\\
		\mathcal{H} &= \left(\frac{\partial \mathcal{L}}{\partial u^{'}}\right)u^{'} - \mathcal{L} = -f(u) \frac{h(u)}{\sqrt{h(u)+u^{'2}}}.
		\label{qaqdistance_Hamiltonian}
	\end{align}
	Following the boundary conditions,
	\begin{equation}
		u(x=\pm \frac{L}{2}) = 0 ,\, u(x=0)=u_0 ,\,\text{and}\,\, u^{'}(x=0)=0,
	\end{equation}
	we can find the equation of motion for $u(x)$ as,
	\begin{equation}
		u^{'} = \sqrt{h(u) \left(\frac{\sigma^2(u)}{\sigma^2(u_0)}-1\right)},
	\end{equation}
where
\begin{equation}
	\sigma(u) = f(u) \sqrt{h(u)}.
\end{equation}
	Finally the quark-antiquark distance $L$ is obtained as,
	\begin{equation}
		L = \int_{-\frac{L}{2}}^{\frac{L}{2}} dx = 2 \int_{0}^{u_0} \frac{1}{u^{'}}du =  2 \int_{0}^{u_0} du\left[h(u)\left(\frac{f(u)^2 h(u)}{f(u_0)^2 h(u_0)}-1\right)\right]^{-\frac{1}{2}}.
	\end{equation}
The quark-antiquark distance $L$ with respect to $u_0$ is plotted in figure (\ref{QAQ_dis_Vs_u0}). Irrespective of the chemical potential, if the temperature is above the small black hole solution region according to the figure (\ref{Temp_Vs_uh_showing_region_of_BH_solution}),  the distance between quark and antiquark increases initially, reaches maximum value and finally breaks down as shown in the red curves in figure (\ref{QAQ_dis_Vs_u0}).  Which correspond to the straight string configuration or unbound state of quark-antiquark pair, representing the deconfinement phase in holographic QCD.
 However, if the temperature is within the small black hole region, there are two types of behaviour depending on the value of chemical potential. For zero chemical potential, blue curve in sub figure (a) of figure (\ref{QAQ_dis_Vs_u0}), the distance between quark antiquark goes to infinity but the tip of the string $u_0$ restricted not to go beyond a certain value along the radial direction of the small black hole spacetime. Therefore the quark-antiquark pair remains intact or U-shape due to the formation of a dynamic wall. This corresponds to the bound state of quark-antiquark pair, representing the confinement phase in holographic QCD. Whereas for non zero chemical potential, the blue curve in sub figure (b) of figure (\ref{QAQ_dis_Vs_u0}), the distance $L$ increases and takes maximum value then decreases slowly and finally it becomes a fixed value at large $u_0$. Which has the similar nature like charged AdS spacetime as shown in the figure (\ref{QAQ_dis_Pure_AdS_and_nonzero_phi}).
\\

From the overall observations, it can be concluded that, for large black hole and three black holes solution zone, the quark-antiquark pair undergoes breakdown forming two separate strings, corresponding to the unbound state and deconfinement phase of the quark-antiquark pair. For small black hole with zero chemical potential, the dynamic wall prevents the breakdown of quark-antiquark pair, corresponding to the bound state and confinement phase. For small black hole with nonzero chemical potential, the quark-antiquark pair doesn't undergo a breakdown, which corresponds to the bound state, though the dynamic wall is not present. Therefore the analysis of boundary theory demands that even in bulk theory there is small black hole instead of AdS space in low temperature, in the boundary theory bound state will be present and as temperature increases the bound state goes to a unbound state through a phase transition called confinement/deconfinement transition.
\\

In the next section we study the Cardy-Verlinde formula for boundary gauge theory.


\section{Cardy-Verlinde Formula for Dual Gauge Theory}\label{cardyverlinde}
According to the AdS/CFT correspondence, the thermodynamical quantity of the AdS bulk spacetime can be identified with the thermodynamical quantities of boundary gauge theory at high temperature. Furthermore, the entropy of the boundary gauge theory can be described by the Cardy-Verlinde formula \cite{Verlinde2000a, Savonije2001,Cai2002b,Cai2001c,Brevik2004,Biswas2004,Biswas2001a,Cappiello2001,Nojiri2002,Nojiri2002,Nojiri2001d,Cai2004}. In this section, we are interested to see whether the entropy of the corresponding boundary theory of our considered spacetime satisfies the Cardy-Verlinde formula or not. In order to verify this we write down the boundary metric from the bulk metric of equation (\ref{metric_ansatz}). Here, the boundary of the bulk geometry represents a hypersurface at a large $r$ distance. Therefore, the boundary metric can be written as,
\begin{equation}
\label{boundarymetric}
ds^2 = \frac{r^2}{l^2}\big[-dt^2 + l^2 d\Omega_{n-1}^2\big].
\end{equation}
By scaling the time parameter as $ \tau = \frac{R}{l}t$ the boundary metric recast as,
\begin{equation}
\label{boundarymetric1}
ds^2 = \frac{r^2}{R^2}\big[-d\tau^2 + R^2 d\Omega_{n-1}^2\big].
\end{equation}
Which is conformally equivalent to $S^{n-1} \times \cal R$ with $S^{n-1}$ having radius of $R$. Using the AdS/CFT correspondence the thermodynamical quantities of the boundary theory can be computed. Certain quantities such as entropy, charge, and string density remain unchanged when mapped from the bulk spacetime to the boundary theory. However, other quantities such as energy, temperature, pressure, chemical potential, Landau function, and the conjugate of the string density ($\mathcal{A}^a$)  can be modified in the mapping process as,
\begin{equation}
	\label{boundarymapping}
	\tilde E = \tilde M + \tilde E_0 = \frac{l}{R} E, \quad \tilde T  = \frac{l}{R}T,\quad \tilde P  = \frac{l}{R} P, \quad \tilde \Phi  = \frac{l}{R}\Phi, \quad \tilde G  = \frac{l}{R}G \,\, {\rm and }\,\,{\cal \tilde A}^a = \frac{l}{R} {\cal A}^a.
\end{equation}
  Here, the conjugate of string density takes the value $ {\cal A}^a = \frac{\partial E}{\partial a} = \frac{\partial M}{\partial a}= -\omega_{n-1} r_+$  for $E_0$ independent of string density $a$ as derived in the equation (\ref{rvacezero}). Comparing with the chemical potential we can consider $\mathcal{A}^a$ as the amount of energy transfer to the background by the strings. Then the generalised first law of thermodynamics of the boundary gauge theory can be expressed as,
\begin{equation}
	\label{boundarythermodynamics}
	d\tilde E =  d\tilde M = \tilde T dS - \tilde PdV +\tilde\Phi dQ +{\cal \tilde A}^a da.
\end{equation}
Here, we consider $r_{\rm vac}$ is fixed and $\tilde P$ is the pressure of the boundary gauge theory and it takes the value
\begin{equation}
\tilde P = -\left(\frac{\partial \tilde E}{\partial V}\right)_{S,\, Q,\, a}  = -\left(\frac{\partial \tilde M}{\partial V}\right)_{S,\, Q,\, a}   = \frac{\tilde M}{(n-1) \omega_{n-1} R^{n-1}}  = \frac{\tilde M}{(n-1) V}.
\end{equation}
Similarly other intensive parameters can be calculated as 
\begin{equation}
	\tilde{T} = \left.\left(\frac{\partial \tilde{M}}{\partial S}\right)\right|_{V,Q,a},
\end{equation}
\begin{equation}
	\tilde{\phi} = \left.\left(\frac{\partial \tilde{M}}{\partial Q}\right)\right|_{S,V,a}.
\end{equation}
Therefore, we can write the Euler equation as \cite{Verlinde2000a}
,
\begin{equation}
	\tilde{M} = \tilde{T}S -\tilde{P}V + \tilde{\phi} Q + \tilde{\mathcal{A}^a}a,
\end{equation}
which can be reduced as, 
\begin{equation}
	n \tilde{M} - (n-1) (\tilde{T}S + \tilde{\phi} Q + \tilde{\mathcal{A}^a}a) = 0.
	\label{CV_Euler_Equation_reduced}
\end{equation}

\subsection{Cardy-Verlinde Formula from Casimir Energy}
In \cite{Verlinde2000a}, it has been discussed that the energy of the boundary gauge theory is not an extensive quantity. It has two parts;  extensive $(E_E)$ and non-extensive $(E_C)$ which is called the Casimir energy. The total energy can be expressed in terms of extensive and non-extensive energy as,
\begin{equation}
	\label{boundaryenergy}
	\tilde E =  E_E + \frac{1}{2}E_C.
\end{equation}
Following \cite{Verlinde2000a,Cai2004}, the Casimir energy is defined as the violation of the Euler equation (\ref{CV_Euler_Equation_reduced}) as,
\begin{equation}
	\label{casimir_energy_formula}
	E_C = n \tilde M - (n-1)( \tilde T S +  \tilde\Phi Q + \mathcal{\tilde A}^a a).
\end{equation}
Using the above values in equation (\ref{casimir_energy_formula}), we get,
\begin{equation}
	\label{casimirenergyequation}
	E_c = \frac{l}{R} \frac{k(n-1) \omega_{n-1}}{8\pi G_{n+1}} r_+^{n-2}.
\end{equation}
Finally, using (\ref{boundaryenergy}) and (\ref{casimirenergyequation}), we can deduce the entropy in the following form,
\begin{equation}
	S = \frac{4\pi R}{(n-1)\sqrt{2}}\sqrt{\frac{E_c}{k}\left(\tilde{M}-\tilde{E}_Q-\tilde{E}_a - \frac{E_c}{2}\right)}.
\end{equation}
Here, $\tilde E_Q = \frac{\tilde\Phi}{2}Q \propto \frac{1}{r_+} $ is the energy of the electromagnetic field outside the black hole horizon and $\tilde E_a = \mathcal{\tilde A}^a a \propto - r_+  $, is the energy contributed due to the presence of strings hanging from the boundary of the AdS space to the bulk. To investigate the nature of this energy we recall the Cornell potential \cite{Eichten:1978tg}, which matches with the potential energy of the heavy quark potential.
\begin{equation}
V(r) = -\frac{\kappa}{r} + \sigma_s r + C,
\end{equation}
where $ \sigma_s$ is related with the string tension. The first term is related to the Coulomb potential dominates in the short distances and second term related to the binding energy of the quarks dominates at the large distances. Therefore comparing the nature of $ \tilde E_Q $ and $\tilde E_a $ with the Cornell potential we can say that the energy $\tilde E_a $ could be related to the binding energy of the quark anti-quark pair represented by the endpoints of the strings in the boundary CFT.\\

Though the Cardy-Verlinde formula takes the usual form with fixed $E_0$. But for $k = -1 $, the Casimir energy $E_C$ becomes negative,  as observed in \cite{Cai2001b} for the AdS spacetime with hyperbolic curvature. Therefore,  the dual gauge theory can not be unitary as discussed in \cite{Cai2002b,Cai2002c}. We may say that for fixed $E_0$, the Cardy-Verlinde formula is only valid for charged AdS black hole with string cloud for flat and spherical curvature space. If $E_0$ is not constant then entropy can not be formulated as the Cardy-Verlinde formula.

Following \cite{Cappiello2001}, this formula can be rederived from the Landau function of the boundary theory, which has been presented in the next section.


\subsection{Cardy-Verlinde Formula from the Landau Function}
Using the same scaling prescription of equation (\ref{boundarymapping}), we can also find the Landau function for the boundary theory from the expression of bulk Landau function and it takes the form as,
\begin{align}
	\label{landau function final value with n and k }
		\tilde{G}(r_+,\tilde{T})=\frac{\omega_{n-1}}{48 \pi G_{n+1} R} \left[ \frac{3(n-1)}{l} r_+^n - 12 \pi R \tilde{T} r_+^{n-1} + 3 l [k(n-1) - 2 (n-2) \Phi^2] r_+^{n-2} \right.\nonumber\\- \left.48 \pi G_{n+1} l a r_+ \right] + \frac{l}{R}E_0.
\end{align}
The boundary Landau function (\ref{landau function final value with n and k }) can be recasted as
\begin{equation}
	\label{landau function in terms of Ec}
	\tilde{G}(r_+,\tilde{T}) = \frac{1}{2}E_c \left[1- \frac{4 \pi R \tilde{T}}{ k l(n-1)} r_+ + \frac{r_+^2}{k l^2} - \frac{2(n-2)}{k(n-1)} \Phi^2 - \frac{16\pi G_{n+1} a}{k(n-1) r_+^{n-3}}   \right]+ \frac{l}{R}E_0,
\end{equation}
where
\begin{equation}
	\label{Ec from landau function}
	E_c = \frac{l}{R} \frac{k(n-1) \omega_{n-1}}{8\pi G_{n+1}} r_+^{n-2}.
\end{equation}
We also know that $\tilde{G} = \tilde{E} - \tilde{T} S - \tilde{\Phi} Q$, from which we can say that
\begin{equation}
	\label{entropy in terms of Ec}
	S = \frac{2\pi R}{k l(n-1) } E_c r_+.
\end{equation}
Then we can also write,
\begin{equation}
	\label{E-Q phi in terms of Ec}
	\tilde{M} - Q \tilde{\Phi} = \frac{1}{2}E_c \left[1 + \frac{r_+^2}{k l^2} - \frac{2(n-2)}{k(n-1)} \Phi^2 - \frac{16\pi G_{n+1} a}{k(n-1) r_+^{n-3}}   \right].
\end{equation}
Now using (\ref{entropy in terms of Ec}) and (\ref{E-Q phi in terms of Ec}), $S$ can be written in terms of $\tilde{E}$ and $E_c$ as,
\begin{equation}
	S = \frac{2\pi R}{(n-1) \sqrt{k}}\left[ E_c \left( 2\tilde{M} - 2 Q \tilde{\Phi}+ \frac{2(n-2) \Phi^2}{k(n-1)}E_c  + \frac{16\pi G_{n+1} a}{k(n-1)r_+^{n-3}}E_c - E_c\right)  \right]^{\frac{1}{2}}.
\end{equation}
Simplifying the $2^{nd},\, 3^{rd}$ and $4^{th}$ terms inside the parenthesis we get,
\begin{equation}
	\label{cardy varlinde formula with n from landau function}
	S = \frac{4\pi  R}{(n-1) \sqrt{2}} \sqrt{\frac{E_c}{k}\left(\tilde{M} - \tilde{E}_Q - \tilde{E}_a- \frac{E_c}{2}\right)},
\end{equation}
where,
\begin{eqnarray}
	&\tilde{E}_Q  = \frac{Q \tilde{\Phi}}{2} \nonumber\\
	&\tilde{E}_a = a \tilde{\mathcal{A}^a} \nonumber\\
	&\tilde{\mathcal{A}^a} = \frac{l}{R} \mathcal{A}^a  \nonumber\\
	&\mathcal{A}^a = \frac{\partial M}{\partial a} = - \omega_{n-1} r_+.
\end{eqnarray}
The Cardy-Verlinde formula derived from the Landau function is similar to the Cardy-Verlinde formula derived from Casimir energy in the last subsection.\\

Since Cardy-Verlinde formula is valid in the dual gauge theory of charged AdS black hole in presence of string cloud, in the line of discussion of \cite{Cappiello2001, Biswas2004}, we can also think that in this boundary theory the quark and antiquark pair should have the first order phase transition between bound and unbound state at a critical temperature. From equation (\ref{landau function final value with n and k }), we can also show the above transition for different temperatures. But we are here avoiding to repeat the similar study.


\pagebreak
\section{Summary and Discussion}
\label{summary}
In this work, we have studied the charged AdS black hole solutions in the presence of strings cloud. Irrespective of the nature of curvature of spacetime at zero temperature, there is one small black hole that exists. Thermodynamical quantities related to the black hole solutions are computed from the Euclidean action of the system by subtracting the contribution of the small black hole present at zero temperature.  We confirmed that black hole solutions maintained the first law of thermodynamics. For flat and hyperbolic nature of the AdS space, there exist only one black hole solution for any temperature, chemical potential and string density. For spherical curvature of the AdS space, large chemical potential and large string density again one black hole solution admits. However, for small enough chemical potential and string density, three black hole solutions exist in a certain temperature range. Beyond the temperature range, only one black hole solution exists. Amongst the three, the medium sized black hole is unstable while the small and large sized  black holes are stable. Depending on the temperature one is thermodynamically more favoured than the other. The Hawking-Page like phase transition takes place between these two black holes at a critical temperature. Since AdS space is no longer a valid phase in this system, one can expect that in the corresponding dual gauge theory there might not be any confined state of quarks and antiquarks. To explore further, we map some of the bulk results into the boundary theory and study the existence of bound state of quark-antiquark pair through the breakdowns of connecting string between quark-antiquark pair and  establish the validity of the Cardy-Verlinde formula. Both the study indicate the presence of bound state in the boundary theory at low temperature. We also observe an additional energy term arising in the Cardy-Verlinde formula
 because of the presence of string cloud.


\section{Acknowledgement}
We would like to thank S. Mukhopadhyay for the insightful discussions and S. Mukherji for going through the manuscript and providing the valuable comments on our work. RP thanks A. Lobo and B. Sharma for fruitful discussions. We would like to thank the referee for the meticulous review and valuable comments, which led to substantial improvement of the manuscript.

\printbibliography

@article{Cai2002c,
archivePrefix = {arXiv},
arxivId = {hep-th/0111093v2},
author = {Cai, Rong Gen},
doi = {10.1016/S0370-2693(01)01457-5},
eprint = {0111093v2},
issn = {03702693},
journal = {Phys. Lett. Sect. B Nucl. Elem. Part. High-Energy Phys.},
month = {nov},
number = {3-4},
pages = {331--336},
primaryClass = {hep-th},
title = {{Cardy-Verlinde formula and asymptotically de Sitter spaces}},
url = {http://arxiv.org/abs/hep-th/0111093 http://dx.doi.org/10.1016/S0370-2693(01)01457-5},
volume = {525},
year = {2002}
}

@article{Headrick:2007ca,
    author = "Headrick, Matthew",
    title = "{Hedgehog black holes and the Polyakov loop at strong coupling}",
    eprint = "0712.4155",
    archivePrefix = "arXiv",
    primaryClass = "hep-th",
    reportNumber = "SU-ITP-07-23",
    doi = "10.1103/PhysRevD.77.105017",
    journal = "Phys. Rev. D",
    volume = "77",
    pages = "105017",
    year = "2008"
}

@article{Bigazzi:2011it,
    author = "Bigazzi, Francesco and Cotrone, Aldo L. and Mas, Javier and Mayerson, Daniel and Tarrio, Javier",
    title = "{D3-D7 Quark-Gluon Plasmas at Finite Baryon Density}",
    eprint = "1101.3560",
    archivePrefix = "arXiv",
    primaryClass = "hep-th",
    reportNumber = "ITP-UU-10-39, DFTT-27-2010",
    doi = "10.1007/JHEP04(2011)060",
    journal = "JHEP",
    volume = "04",
    pages = "060",
    year = "2011"
}

@article{Kumar:2012ui,
    author = "Kumar, S. Prem",
    title = "{Heavy quark density in N=4 SYM: from hedgehog to Lifshitz spacetimes}",
    eprint = "1206.5140",
    archivePrefix = "arXiv",
    primaryClass = "hep-th",
    doi = "10.1007/JHEP08(2012)155",
    journal = "JHEP",
    volume = "08",
    pages = "155",
    year = "2012"
}

@article{Cardy:1986ie,
    author = "Cardy, John L.",
    title = "{Operator Content of Two-Dimensional Conformally Invariant Theories}",
    doi = "10.1016/0550-3213(86)90552-3",
    journal = "Nucl. Phys. B",
    volume = "270",
    pages = "186--204",
    year = "1986"
}

@article{Dey:2020yzl,
    author = "Dey, Tanay K. and Mukhopadhyay, Subir",
    title = "{AdS black holes with higher derivative corrections in presence of string cloud}",
    eprint = "2005.12054",
    archivePrefix = "arXiv",
    primaryClass = "hep-th",
    doi = "10.1140/epjc/s10052-020-08565-9",
    journal = "Eur. Phys. J. C",
    volume = "80",
    number = "11",
    pages = "1012",
    year = "2020"
}

@article{Dey:2023inw,
    author = "Dey, Tanay K. and Mukhopadhyay, Subir",
    title = "{Charged AdS black holes with higher derivative corrections in presence of string cloud}",
    eprint = "2302.00256",
    archivePrefix = "arXiv",
    primaryClass = "hep-th",
    month = "2",
    year = "2023"
}

@article{Brevik2004,
archivePrefix = {arXiv},
arxivId = {hep-th/0401073},
author = {Brevik, Iver and Nojiri, Shin'ichi and Odintsov, Sergei D. and Vanzo, Luciano},
doi = {10.1103/PhysRevD.70.043520},
eprint = {0401073},
issn = {15502368},
journal = {Phys. Rev. D - Part. Fields, Gravit. Cosmol.},
number = {4},
primaryClass = {hep-th},
title = {{Entropy and universality of the Cardy-Verlinde formula in a dark energy universe}},
volume = {70},
year = {2004}
}

@article{Verlinde2000a,
archivePrefix = {arXiv},
arxivId = {hep-th/0008140},
author = {Verlinde, Erik},
eprint = {0008140},
number = {August},
primaryClass = {hep-th},
title = {{On the Holographic Principle in a Radiation Dominated Universe}},
url = {http://arxiv.org/abs/hep-th/0008140},
year = {2000}
}

@article{Nojiri2002,
archivePrefix = {arXiv},
arxivId = {hep-th/0205187},
author = {Nojiri, Shin'Ichi I. and Odintsov, Sergei D. and Ogushi, Sachiko},
doi = {10.1142/S0217751X02012156},
eprint = {0205187},
issn = {0217751X},
journal = {Int. J. Mod. Phys. A},
number = {32},
pages = {4809--4870},
primaryClass = {hep-th},
title = {{Friedmann-Robertson-Walker brane cosmological equations from the five-dimensional bulk (A)dS black hole}},
volume = {17},
year = {2002}
}

@article{Liang2021,
archivePrefix = {arXiv},
arxivId = {2008.09512},
author = {Liang, Jing and Mu, Benrong and Tao, Jun},
doi = {10.1088/1674-1137/abd085},
eprint = {2008.09512},
issn = {16741137},
journal = {Chinese Phys. C},
number = {2},
pages = {1--26},
title = {{Thermodynamics and overcharging problem in extended phase space of charged AdS black holes with cloud of strings and quintessence under charged particle absorption}},
volume = {45},
year = {2021}
}

@article{Biswas2004,
archivePrefix = {arXiv},
arxivId = {hep-th/0310238},
author = {Biswas, Anindya and Mukherji, Sudipta},
doi = {10.1016/j.physletb.2003.10.089},
eprint = {0310238},
issn = {03702693},
journal = {Phys. Lett. Sect. B Nucl. Elem. Part. High-Energy Phys.},
number = {3-4},
pages = {425--431},
primaryClass = {hep-th},
title = {{On the Hawking-Page transition and the Cardy-Verlinde formula}},
volume = {578},
year = {2004}
}

@article{Witten2014,
archivePrefix = {arXiv},
arxivId = {hep-th/9803131},
author = {Witten, Edward},
doi = {10.1142/9789814571616_0023},
eprint = {9803131},
isbn = {9789814571616},
journal = {Oskar Klein Meml. Lect. 1988-1999},
pages = {389--419},
primaryClass = {hep-th},
title = {{Anti-de Sitter space, thermal phase transition, and confinement in gauge theories}},
year = {2014}
}

@article{Chakrabortty2011a,
archivePrefix = {arXiv},
arxivId = {1108.0165},
author = {Chakrabortty, Shankhadeep},
doi = {10.1016/j.physletb.2011.09.112},
eprint = {1108.0165},
issn = {03702693},
journal = {Phys. Lett. Sect. B Nucl. Elem. Part. High-Energy Phys.},
number = {3},
pages = {244--250},
title = {{Dissipative force on an external quark in heavy quark cloud}},
volume = {705},
year = {2011}
}

@article{Karch2002,
archivePrefix = {arXiv},
arxivId = {hep-th/0205236},
author = {Karch, Andreas and Katz, Emanuel},
doi = {10.1088/1126-6708/2002/06/043},
eprint = {0205236},
issn = {10298479},
journal = {J. High Energy Phys.},
number = {6},
pages = {1021--1036},
primaryClass = {hep-th},
title = {{Adding flavor to AdS/CFT}},
volume = {6},
year = {2002}
}

@article{Cai2001c,
archivePrefix = {arXiv},
arxivId = {hep-th/0105070},
author = {Cai, Rong Gen and Myung, Yun Soo and Ohta, Nobuyoshi},
doi = {10.1088/0264-9381/18/24/308},
eprint = {0105070},
issn = {02649381},
journal = {Class. Quantum Gravity},
number = {24},
pages = {5429--5440},
primaryClass = {hep-th},
title = {{Bekenstein bound, holography and brane cosmology in charged black hole backgrounds}},
volume = {18},
year = {2001}
}

@article{Witten1998,
archivePrefix = {arXiv},
arxivId = {hep-th/9802150},
author = {Witten, Edward},
doi = {10.4310/atmp.1998.v2.n2.a2},
eprint = {9802150},
issn = {10950753},
journal = {Adv. Theor. Math. Phys.},
number = {2},
pages = {253--290},
primaryClass = {hep-th},
title = {{Anti de sitter space and holography}},
volume = {2},
year = {1998}
}

@article{Dey2018,
archivePrefix = {arXiv},
arxivId = {1711.07008},
author = {Dey, Tanay K.},
doi = {10.1142/S0217751X18501932},
eprint = {1711.07008},
issn = {0217751X},
journal = {Int. J. Mod. Phys. A},
month = {nov},
number = {33},
publisher = {World Scientific Publishing Co. Pte Ltd},
title = {{Phase transition of AdS-Schwarzschild black hole and gauge theory dual in the presence of external string cloud}},
volume = {33},
year = {2018}
}

@article{Ghanaatian2019a,
archivePrefix = {arXiv},
arxivId = {1906.00369},
author = {Ghanaatian, M. and Sadeghi, Mehdi and Ranjbari, Hadi and Forozani, Gh},
eprint = {1906.00369},
journal = {arXiv},
pages = {1--23},
title = {{Effects of the external string cloud on the Van der Waals like behaviour and efficiency of AdS-Schwarzschild black holes in massive gravity}},
year = {2019}
}

@article{Chabab2020,
archivePrefix = {arXiv},
arxivId = {2001.06063},
author = {Chabab, M. and Iraoui, S.},
doi = {10.1007/s10714-020-02729-4},
eprint = {2001.06063},
issn = {15729532},
journal = {Gen. Relativ. Gravit.},
number = {8},
title = {{Thermodynamic criticality of d-dimensional charged AdS black holes surrounded by quintessence with a cloud of strings background}},
volume = {52},
year = {2020}
}

@article{Cai2001b,
archivePrefix = {arXiv},
arxivId = {hep-th/0102113},
author = {Cai, Rong Gen},
doi = {10.1103/PhysRevD.63.124018},
eprint = {0102113},
issn = {05562821},
journal = {Phys. Rev. D},
number = {12},
primaryClass = {hep-th},
title = {{Cardy-Verlinde formula and AdS black holes}},
volume = {63},
year = {2001}
}

@article{Herscovich2010a,
archivePrefix = {arXiv},
arxivId = {1004.3754},
author = {Herscovich, Estanislao and Richarte, Mart{\'{i}}n G.},
doi = {10.1016/j.physletb.2010.04.065},
eprint = {1004.3754},
issn = {03702693},
journal = {Phys. Lett. Sect. B Nucl. Elem. Part. High-Energy Phys.},
number = {4-5},
pages = {192--200},
title = {{Black holes in Einstein-Gauss-Bonnet gravity with a string cloud background}},
volume = {689},
year = {2010}
}

@article{Hawking1983,
author = {Hawking, S. W. and Page, Don N.},
doi = {10.1007/BF01208266},
issn = {00103616},
journal = {Commun. Math. Phys.},
number = {4},
pages = {577--588},
title = {{Thermodynamics of black holes in anti-de Sitter space}},
volume = {87},
year = {1983}
}

@article{Yin2021a,
archivePrefix = {arXiv},
arxivId = {2103.08162},
author = {Yin, Rui and Liang, Jing and Mu, Benrong},
doi = {10.1016/j.dark.2021.100831},
eprint = {2103.08162},
issn = {22126864},
journal = {Phys. Dark Universe},
title = {{Stability of horizon with pressure and volume of d-dimensional charged AdS black holes with cloud of strings and quintessence}},
volume = {32},
year = {2021}
}

@article{Maldacena1999,
archivePrefix = {arXiv},
arxivId = {hep-th/9711200},
author = {Maldacena, Juan},
doi = {10.1023/A:1026654312961},
eprint = {9711200},
issn = {00207748},
journal = {Int. J. Theor. Phys.},
number = {4},
pages = {1113--1133},
primaryClass = {hep-th},
title = {{The large-N limit of superconformal field theories and supergravity}},
volume = {38},
year = {1999}
}

@article{Cai2002b,
archivePrefix = {arXiv},
arxivId = {hep-th/0112253},
author = {Cai, Rong Gen},
doi = {10.1016/S0550-3213(02)00064-0},
eprint = {0112253},
issn = {05503213},
journal = {Nucl. Phys. B},
number = {1-2},
pages = {375--386},
primaryClass = {hep-th},
title = {{Cardy-Verlinde formula and thermodynamics of black holes in de Sitter spaces}},
volume = {628},
year = {2002}
}

@article{Cai2004,
archivePrefix = {arXiv},
arxivId = {hep-th/0410158},
author = {Cai, Rong Gen and Pang, Da Wei and Wang, Anzhong},
doi = {10.1103/PhysRevD.70.124034},
eprint = {0410158},
issn = {15502368},
journal = {Phys. Rev. D - Part. Fields, Gravit. Cosmol.},
number = {12},
pages = {9},
primaryClass = {hep-th},
title = {{Born-Infeld black holes in (A)dS spaces}},
volume = {70},
year = {2004}
}

@article{Cappiello2001,
archivePrefix = {arXiv},
arxivId = {hep-th/0107238},
author = {Cappiello, L. and M{\"{u}}ck, W.},
doi = {10.1016/S0370-2693(01)01283-7},
eprint = {0107238},
issn = {03702693},
journal = {Phys. Lett. Sect. B Nucl. Elem. Part. High-Energy Phys.},
number = {1-2},
pages = {139--144},
primaryClass = {hep-th},
title = {{On the phase transition of conformal field theories with holographic duals}},
volume = {522},
year = {2001}
}

@article{Biswas2001a,
archivePrefix = {arXiv},
arxivId = {hep-th/0102138},
author = {Biswas, Anindya K. and Mukherji, Sudipta},
doi = {10.1088/1126-6708/2001/03/046},
eprint = {0102138},
issn = {10298479},
journal = {J. High Energy Phys.},
number = {3},
pages = {3--8},
primaryClass = {hep-th},
title = {{Holography and stiff-matter on the brane}},
volume = {5},
year = {2001}
}

@article{Anninos2009,
archivePrefix = {arXiv},
arxivId = {0807.3478},
author = {Anninos, Dionysios and Pastras, Georgios},
doi = {10.1088/1126-6708/2009/07/030},
eprint = {0807.3478},
issn = {11266708},
journal = {J. High Energy Phys.},
number = {7},
title = {{Thermodynamics of the Maxwell-gauss-bonnet anti-de Sitter black hole with higher derivative gauge corrections}},
volume = {2009},
year = {2009}
}

@article{Chamblin1999a,
archivePrefix = {arXiv},
arxivId = {hep-th/9902170},
author = {Chamblin, Andrew and Emparan, Roberto and Johnson, Clifford V. and Myers, Robert C.},
doi = {10.1103/PhysRevD.60.064018},
eprint = {9902170},
issn = {15502368},
journal = {Phys. Rev. D - Part. Fields, Gravit. Cosmol.},
number = {6},
primaryClass = {hep-th},
title = {{Charged AdS black holes and catastrophic holography}},
volume = {60},
year = {1999}
}

@article{Nojiri2001d,
archivePrefix = {arXiv},
arxivId = {hep-th/0105117},
author = {Nojiri, Shin'ichi and Odintsov, Sergei D. and Ogushi, Sachiko},
doi = {10.1142/S0217751X01005584},
eprint = {0105117},
issn = {0217751X},
journal = {Int. J. Mod. Phys. A},
number = {31},
pages = {5085--5099},
primaryClass = {hep-th},
title = {{Holographic entropy and brane FRW dynamics from AdS black hole in d5 higher derivative gravity}},
volume = {16},
year = {2001}
}

@article{Savonije2001,
archivePrefix = {arXiv},
arxivId = {hep-th/0102042},
author = {Savonije, Ivo and Verlinde, Erik},
doi = {10.1016/S0370-2693(01)00467-1},
eprint = {0102042},
issn = {03702693},
journal = {Phys. Lett. Sect. B Nucl. Elem. Part. High-Energy Phys.},
number = {1-4},
pages = {305--311},
primaryClass = {hep-th},
title = {{CFT and entropy on the brane}},
volume = {507},
year = {2001}
}

@article{Ghaffarnejad2018a,
archivePrefix = {arXiv},
arxivId = {1806.06687},
author = {Ghaffarnejad, Hossein and Yaraie, Emad},
doi = {10.1016/j.physletb.2018.08.017},
eprint = {1806.06687},
issn = {03702693},
journal = {Phys. Lett. Sect. B Nucl. Elem. Part. High-Energy Phys.},
pages = {105--111},
title = {{Effects of a cloud of strings on the extended phase space of Einstein–Gauss–Bonnet AdS black holes}},
volume = {785},
year = {2018}
}

@article{Balasubramanian1999,

archivePrefix = {arXiv},

arxivId = {hep-th/9902121},

author = {Balasubramanian, Vijay and Kraus, Per},

doi = {10.1007/s002200050764},

eprint = {9902121},

issn = {00103616},

journal = {Commun. Math. Phys.},

number = {2},

pages = {413--428},

primaryClass = {hep-th},

title = {{A stress tensor for anti-de Sitter gravity}},

volume = {208},

year = {1999}

}

@article{Eichten:1978tg,
    author = "Eichten, E. and Gottfried, K. and Kinoshita, T. and Lane, K. D. and Yan, Tung-Mow",
    title = "{Charmonium: The Model}",
    reportNumber = "CLNS-375",
    doi = "10.1103/PhysRevD.17.3090",
    journal = "Phys. Rev. D",
    volume = "17",
    pages = "3090",
    year = "1978",
    note = "[Erratum: Phys.Rev.D 21, 313 (1980)]"
}

@article{Vanzo1997,
	doi = {10.1103/physrevd.56.6475},

	url = {https://doi.org/10.1103%2Fphysrevd.56.6475},

	year = 1997,
	month = {nov},

	publisher = {American Physical Society ({APS})},

	volume = {56},

	number = {10},

	pages = {6475--6483},

	author = {L. Vanzo},

	title = {Black holes with unusual topology},

	journal = {Physical Review D}
}
\end{document}